\documentclass{article}

\usepackage{arxiv}

\usepackage[utf8]{inputenc} 
\usepackage[T1]{fontenc}    
\usepackage{booktabs}       
\usepackage{amsfonts}       
\usepackage{nicefrac}       
\usepackage{microtype}      
\usepackage{lipsum}		
\usepackage{graphicx}
\usepackage[square,numbers]{natbib}
\usepackage{doi}

\usepackage{subfig}
\usepackage{pgfplots} 

\usepackage{acronym}
\usepackage{color, soul}

\usepackage[
  toc,
  acronym,
  style = long
]{glossaries}
\newglossary[slg]{reliability}{syi}{syg}{Nomenclature}

\makeglossaries

\newglossaryentry{symb:E}{
    name=\ensuremath{E},
    description={Set of Electronic Control Units},
    type=reliability}

\newglossaryentry{symb:e}{
    name=\ensuremath{e},
    description={Electronic Control Unit},
    type=reliability}
    
\newglossaryentry{symb:A}{
    name=\ensuremath{A},
    description={Set of applications},
    type=reliability}

\newglossaryentry{symb:a}{
    name=\ensuremath{a},
    description={Application},
    type=reliability}
    
\newglossaryentry{symb:G_a}{
    name=\ensuremath{G_a},
    description={Application instance graph},
    type=reliability}
    
\newglossaryentry{symb:vertices}{
    name=\ensuremath{\mathcal{V}},
    description={Application graph vertices},
    type=reliability}
    
\newglossaryentry{symb:edges}{
    name=\ensuremath{\mathcal{E}},
    description={Application graph edges},
    type=reliability}
    
\newglossaryentry{symb:T}{
    name=\ensuremath{T},
    description={Set of all task instances},
    type=reliability}
    
\newglossaryentry{symb:T_a}{
    name=\ensuremath{T_a},
    description={Set of active task instances},
    type=reliability}

\newglossaryentry{symb:T_p}{
    name=\ensuremath{T_p},
    description={Set of passive task instances},
    type=reliability}

\newglossaryentry{symb:T_c}{
    name=\ensuremath{T_c},
    description={Set of critical task instances},
    type=reliability}

\newglossaryentry{symb:t}{
    name=\ensuremath{t},
    description={A task instance},
    type=reliability}
    
\newglossaryentry{symb:M_a}{
    name=\ensuremath{M_a},
    description={Set of active message instances},
    type=reliability}
    
\newglossaryentry{symb:M_b}{
    name=\ensuremath{M_b},
    description={Set of backup message instances},
    type=reliability}

\newglossaryentry{symb:m}{
    name=\ensuremath{m},
    description={A message instance},
    type=reliability}

\newglossaryentry{symb:L}{
    name=\ensuremath{L},
    description={Set of Ethernet links},
    type=reliability}

\newglossaryentry{symb:l}{
    name=\ensuremath{l},
    description={Ethernet link},
    type=reliability}

\newglossaryentry{symb:P_a}{
    name=\ensuremath{P_a},
    description={Application period},
    type=reliability}
    
\newglossaryentry{symb:delta}{
    name=\ensuremath{\delta},
    description={Application deadline},
    type=reliability}
    
\newglossaryentry{symb:W}{
    name=\ensuremath{W},
    description={Worst-case execution time},
    type=reliability}

\newglossaryentry{symb:alpha}{
    name=\ensuremath{\alpha},
    description={Binding function of active task instances},
    type=reliability}

\newglossaryentry{symb:beta}{
    name=\ensuremath{\beta},
    description={Binding function of passive task instances},
    type=reliability}

\newglossaryentry{symb:rho}{
    name=\ensuremath{\rho},
    description={Routing function of active message instances},
    type=reliability}
    
\newglossaryentry{symb:sigma}{
    name=\ensuremath{\sigma},
    description={Routing function of backup message instances},
    type=reliability}

\newglossaryentry{symb:SI_{max}}{
    name=\ensuremath{SI_{max}},
    description={Maximum number of available slots},
    type=reliability}

\newglossaryentry{symb:SI_r}{
    name=\ensuremath{SI_r},
    description={Set of slots reserved for a passive task instance},
    type=reliability}

\newglossaryentry{symb:SI_a}{
    name=\ensuremath{SI_a},
    description={Set of slots reserved for an active task instance},
    type=reliability}
\glsadd{symb:SI_a}

\newglossaryentry{symb:N_{SI,a}}{
    name=\ensuremath{N_{SI,a}},
    description={Total number of slots on an ECU that are exclusively allocated},
    type=reliability}
    
\newglossaryentry{symb:N_{SI,r}}{
    name=\ensuremath{N_{SI,r}},
    description={Total number of slots on an ECU that are exclusively reserved},
    type=reliability}

\newglossaryentry{symb:N_{SI,ar}}{
    name=\ensuremath{N_{SI,ar}},
    description={Total number of slots on an ECU that are both allocated and reserved},
    type=reliability}

\newglossaryentry{symb:N_{SL,a}}{
    name=\ensuremath{N_{SL,a}},
    description={Total number of slots on a link that are exclusively allocated},
    type=reliability}

\newglossaryentry{symb:N_{SL,r}}{
    name=\ensuremath{N_{SL,r}},
    description={Total number of slots on a link that are exclusively reserved},
    type=reliability}

\newglossaryentry{symb:SI}{
    name=\ensuremath{SI},
    description={A scheduling slot},
    type=reliability}

\newglossaryentry{symb:B}{
    name=\ensuremath{B},
    description={The search space of a passive task instance},
    type=reliability}

\newglossaryentry{symb:tau}{
    name=\ensuremath{\tau},
    description={A specified amount of time},
    type=reliability}

\newglossaryentry{symb:R}{
    name=\ensuremath{R},
    description={Reliability},
    type=reliability}

\newglossaryentry{symb:lambda}{
    name=\ensuremath{\lambda},
    description={Failure rate},
    type=reliability}
    
\newglossaryentry{symb:MTTF}{
    name=\ensuremath{MTTF},
    description={Mean Time to Failure},
    type=reliability}
    
\newglossaryentry{symb:varphi}{
    name=\ensuremath{\varphi},
    description={Structure function},
    type=reliability}

\newglossaryentry{symb:specialset}{
    name=\ensuremath{T_{t_a}^r},
    description={Set of tasks of critical applications where a passive task instance has reserved a slot that is allocated by the task instance $t_a$},
    type=reliability}

\newglossaryentry{symb:MTTF_{AVG}}{
    name=\ensuremath{MTTF_{AVG}},
    description={The averaged Mean Time to Failure},
    type=reliability}
    
\newglossaryentry{symb:S_e}{
    name=\ensuremath{S_e},
    description={Slot capacity of an ECU},
    type=reliability}

\newglossaryentry{symb:S_t}{
    name=\ensuremath{S_t},
    description={Total slot capacity of the system},
    type=reliability}
    
\newglossaryentry{symb:N_a}{
    name=\ensuremath{N_a},
    description={Total amount of applications},
    type=reliability}

\newglossaryentry{symb:N_c}{
    name=\ensuremath{N_c},
    description={Total amount of critical applications},
    type=reliability}

\newglossaryentry{symb:N_{nc}}{
    name=\ensuremath{N_{nc}},
    description={Total amount of non-critical applications},
    type=reliability}

\newglossaryentry{symb:S_O}{
    name=\ensuremath{S_O},
    description={Total amount of occupied slots in a system},
    type=reliability}
    
\newglossaryentry{symb:MTTF_{reduction,nc}}{
    name=\ensuremath{MTTF_{reduction,nc}},
    description={Percental MTTF reduction of non-critical applications of our degradation approach compared to the active redundancy approach},
    type=reliability}
    
\newglossaryentry{symb:MTTF_{AVG,active,nc}}{
    name=\ensuremath{MTTF_{AVG,active,nc}},
    description={Average MTTF of non-critical applications of the active redundancy approach},
    type=reliability}
    
\newglossaryentry{symb:MTTF_{AVG,deg,nc}}{
    name=\ensuremath{MTTF_{AVG,deg,nc}},
    description={Average MTTF of non-critical applications of our graceful degradation approach},
    type=reliability}
    
\newglossaryentry{symb:S_{OH,deg}}{
    name=\ensuremath{S_{OH,deg}},
    description={Slot overhead introduced by our degradation approach},
    type=reliability}
    
\newglossaryentry{symb:S_{O,deg}}{
    name=\ensuremath{S_{O,deg}},
    description={Total number of consumed slots of our degradation approach},
    type=reliability}
    
\newglossaryentry{symb:S_{O,no}}{
    name=\ensuremath{S_{O,no}},
    description={Total number of consumed slots when using no redundancy},
    type=reliability}

\newglossaryentry{symb:S_{OH,active}}{
    name=\ensuremath{S_{OH,active}},
    description={Slot overhead introduced by the active redundancy approach},
    type=reliability}
    
\newglossaryentry{symb:S_{O,active}}{
    name=\ensuremath{S_{O,active}},
    description={Total number of consumed slots of the active redundancy approach},
    type=reliability}
    
\newglossaryentry{symb:R_{savings}}{
    name=\ensuremath{R_{savings}},
    description={Percental resource savings of our degradation approach over the active redundancy approach},
    type=reliability}

\newglossaryentry{symb:p}{
    name=\ensuremath{p},
    description={Probability function},
    type=reliability}

\newglossaryentry{symb:f}{
    name=\ensuremath{f},
    description={A boolean function},
    type=reliability}

\newglossaryentry{symb:x}{
    name=\ensuremath{x},
    description={A boolean variable},
    type=reliability}

\newglossaryentry{symb:z}{
    name=\ensuremath{z},
    description={Function which indicates the operational status of a task instance},
    type=reliability}

\newglossaryentry{symb:y}{
    name=\ensuremath{y},
    description={Function which indicates the operational status of an ECU},
    type=reliability}

\newglossaryentry{symb:u}{
    name=\ensuremath{u},
    description={Function which translates a boolean variable z which is dependent on the state of a task instance to a boolean variable y which is dependent on the state of an ECU},
    type=reliability}

\newglossaryentry{symb:P}{
    name=\ensuremath{P},
    description={Probability of a proper working system},
    type=reliability}

\newcommand{\markasnew}[1]{{#1}}	  
\newcommand{\marknew}[1]{{#1}}	  

\title{Reliability Analysis of Gracefully Degrading Automotive Systems}

\author{ \href{https://orcid.org/0000-0000-0000-0000}{\includegraphics[scale=0.06]{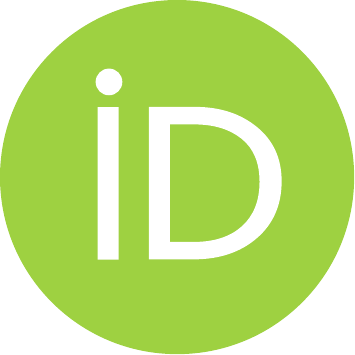}\hspace{1mm}Philipp Weiss} \\
	Technical University of Munich\\
	80333 Munich, Germany \\
	\texttt{philipp.weiss@tum.de} \\
    \And
	\href{https://orcid.org/0000-0000-0000-0000}{\includegraphics[scale=0.06]{orcid.pdf}\hspace{1mm}Ali Younessi} \\
	Technical University of Munich\\
	80333 Munich, Germany \\
	\texttt{ali.younessi@tum.de} \\
	\And
	\href{https://orcid.org/0000-0000-0000-0000}{\includegraphics[scale=0.06]{orcid.pdf}\hspace{1mm}Sebastian Steinhorst} \\
	Technical University of Munich\\
	80333 Munich, Germany \\
	\texttt{sebastian.steinhorst@tum.de} \\
}

\renewcommand{\shorttitle}{\textit{arXiv} Template}

\hypersetup{
pdftitle={A template for the arxiv style},
pdfsubject={q-bio.NC, q-bio.QM},
pdfauthor={David S.~Hippocampus, Elias D.~Striatum},
pdfkeywords={First keyword, Second keyword, More},
}

\begin{document}
\maketitle

\begin{abstract}
Fail-operational systems are a prerequisite for autonomous driving.
Without a driver who can act as a fallback solution in a critical failure scenario, the system has to be able to mitigate failures on its own and keep critical applications operational.
To reduce redundancy cost, graceful degradation can be applied by repurposing hardware resources at run-time. 
Critical applications can be kept operational by starting passive backups and shutting down non-critical applications instead to make sufficient resources available.
In order to design such systems efficiently, the degradation effects on reliability and cost savings have to be analyzed.

In this paper we present our approach to formally analyze the impact of graceful degradation on the reliability of critical and non-critical applications.
We then quantify the effect of graceful degradation on the reliability of both critical and non-critical applications in distributed automotive systems and compare the achieved cost reduction with conventional redundancy approaches.

In our experiments redundancy overhead could be reduced by $80 \%$ compared to active redundancy in a scenario with a balanced mix of critical and non-critical applications using our graceful degradation approach.
Our results show that a trade-off between the impact of the degradation on the reliability of non-critical applications and cost reduction has to be made. 

Overall, we present a detailed reliability and cost analysis of graceful degradation in distributed automotive systems.
Our findings confirm that using graceful degradation can tremendously reduce cost compared to conventional redundancy approaches with no negative impact on the redundancy of critical applications if a reliability reduction of non-critical applications can be accepted.
\end{abstract}


\section{Introduction}

To enable autonomous driving a fail-operational behavior of automotive systems is essential. 
As there is no driver available as a backup solution in case of the failure of an Electronic Control Unit (ECU), the software system must be able to mitigate failures itself.
Automotive architectures are currently undergoing major changes to manage increased software and communication complexity.
Software is being integrated on a few, more powerful central control units.
Additionally, the system must be able to integrate new functionalities via over-the-air software updates.
Here, customers will be able to purchase and enable features over an app store, which leads to unique and customized software systems.

To handle frequent software updates and increased safety demands a predictable and dynamic resource management is required.
Such a dynamic resource management allows to integrate new applications at run-time with unique software solutions and enables a safe behavior of critical applications at the same time.
To reduce hardware resource consumption and costs, a gracefully degrading behavior can be implemented in such a dynamic resource management.
With a graceful degradation approach safety-critical tasks can be restarted on other available hardware resources, while, in return, non-critical tasks are
shut down to free resources. 
Instead of adding costly hardware redundancy to enable a fail-operational
behaviour, existing resources can be repurposed. 
Compared to an active redundancy approach, no additional hardware resources are required as resources for the backup tasks are shared with non-critical tasks.
Decentralized run-time approaches have the advantage that there is
no single-point-of-failure such that the system is still able to act after any ECU failure \cite{WWRS20}.

The main challenges for such a system are to achieve a predictable behavior.
As resources are shifted dynamically, non-critical applications that are not directly affected by an ECU failure might get shut down to free resources for restarting critical tasks.
Thus, with a graceful degradation approach, the reliability of critical applications is increased at the cost of a reliability decrease of non-critical applications.
Here, it is important to quantify and understand the effect that a graceful degradation approach has on both critical and non-critical applications to increase predictability of this approach.

In previous work in \cite{weiss2023predictable} \markasnew{we have presented such a gracefully degrading system architecture.
The presented scheduling approach supports a gracefully degrading behaviour such that resources of non-critical applications might be taken from re-starting critical applications.
The applications are mapped onto the architecture using an agent-based approach. 
The focus of the work in} \cite{weiss2023predictable} \markasnew{was the predictable timing analysis of the applications. 
However, there are remaining questions regarding the impact of the graceful degradation approach on the reliability of the system as the reliability of critical applications is increased at the cost of a reliability decrease of non-critical applications.
There has not been any work which analyzed the trade-off between additional resource consumption and impact on reliability of graceful degradation or compared it to approaches such as active redundancy.
Therefore, we make the following contributions:}
\begin{itemize}
    \item \markasnew{We introduce our approach to formally analyze the impact of graceful degradation on the reliability of critical and non-critical applications. This approach specifically considers that scheduling slots of non-critical applications can be reserved by critical applications effectively reducing the reliability.}
    \item \markasnew{We perform experiments in our in-house developed simulation framework which simulates the agent-based approach on a virtual architecture.
    Here, we present, for the first time, an in-depth trade-off analysis of a graceful degradation approach where we analyze the resource consumption and the impact of graceful degradation on the reliability of critical and non-critical applications.
    In our experiments we evaluate our three allocation and reservation strategies and compare them to active redundancy.
    Additionally, we present experimental results of our \textit{Predecessor Heuristic} which aims at reducing the exposure of the individual applications to failure sources. }
    \item  \markasnew{We conclude that graceful degradation can be a powerful methodology which reduces resource consumption compared to active redundancy while providing the same reliability to critical applications as an active redundancy approach. However, this is bought with a reduced reliability of non-critical applications.}
\end{itemize}

\markasnew{The outline of our work is as follows:}
\begin{itemize}

    \item \markasnew{We present related work in the fields of fail-operational systems, graceful degradation and reliability analysis in Section} \ref{related_work} \markasnew{and conclude that there is a lack of work on reliability analysis of graceful degradation approaches.}
    \item \markasnew{After presenting our system model in Section} \ref{system_model} we present in Section \ref{graceful} \markasnew{our existing graceful degradation approach from} \cite{weiss2023predictable} \markasnew{which we use to perform our reliability and cost analysis. Our system allows an independent allocation and reservation of resources by each application.} \markasnew{Here, we also introduce three different strategies for the allocation and reservation for resources.}
    \item \markasnew{In Section} \ref{reliability_analysis} \markasnew{we introduce our approach to formally analyze the impact of graceful degradation on the reliability of critical and non-critical applications.}
    \item \markasnew{Last, we perform experiments in our in-house developed simulation framework which simulates the agent-based approach on a virtual architecture in Section} \ref{evaluation}. \markasnew{Here, we analyze the resource consumption and the impact of graceful degradation on the reliability of critical and non-critical applications.}
\end{itemize}

\section{Related Work}
\label{related_work}
The aim of this section is to identify which state-of-the-art approaches are available in literature, categorize them and find their relation to our approach in this paper.
We are first giving an overview over fail-operational systems in Subsection \ref{sec:failop}.
Afterwards we introduce related work on graceful degradation approaches in Subsection \ref{sec:graceful}.
Last, we discuss previous work on the topic of reliability analysis in Subsection \ref{sec:reliability}.

\subsection{Fail-Operational Systems}
\label{sec:failop}
Ensuring fail-operational behavior is crucial in new emerging automotive systems to meet reliability requirements. 
Especially given that the future of automotive vehicles is trending towards autonomous driving with no driver to fall back on. 
Therefore, fail-safe and fail-silent state-of-the-art approaches are no longer sufficient.

In the traditional sense making a system fail-safe is usually done by monitoring,  hardware redundancy (resource replication), or special shutdown procedures \cite{7363617}. 
These solutions have other trade-offs such as cost, area and manufacturing overhead, which are  mainly due to the additional hardware requirements \cite{10.1007/978-3-642-16576-4_4}. 
This method is often known as structural redundancy which applies often to lower abstraction levels such as circuit or device level.
The authors in \cite{1035218} present an overview of different fault tolerant designs that achieve various levels of fail-operational, fail-silent and fail-safe systems based on changing degrees of redundancy for drive-by-wire systems. 
\markasnew{In} \cite{Kohn15}, \markasnew{the authors present an overview on existing fail-operational hardware approaches and introduce concepts for the implementation on a multi-core processor.}
\markasnew{The authors in} \cite{Baleani03} \markasnew{review common fault-tolerant architectures in SoC solutions such as lock-step architectures, loosely synchronized processors or triple modular redundancy and perform a trade-off analysis.
These presented detection and mitigation mechanisms are a base for enabling a dynamic fail-operational solution on software level.
However, the presented approaches miss to apply fail-operational aspects on a system level instead of device level only and are less flexible.
By comparison our system wide graceful degradation approach utilizes the possibility of re-starting applications and using the resources of non-critical applications such that resource overhead introduced by redundancy can be reduced.
 }

A method presented in \cite{Mariani2010AFM} approaches this problem at the hardware architecture level. 
The paper describes a hardware architecture that meets the strict functional safety norms of the standards IEC61508 and ISO26262.
Although it does this by improving costs using a pre-certified hardware fault supervisor, this method only focuses on the hardware level and does not take into account the ever-increasing complexity of functions and applications running on these ECUs.

Approaches such as presented in \cite{Ferrari} and \cite{Delphi} use a dual lock-step architecture where two identical CPUs are used to peform the same software tasks.
The first CPU is in a live mode and controls the system under normal operational mode (no fault present). 
The second CPU checks the status of the first CPU every clock cycle. 
The authors propose a trade-off between performance and fault coverage. 
The system can be used as two fail-silent channels or as a single fail-operational unit. 


 


The authors in \cite{6625234} present a coded processing approach which works by coding data and instructions. 
For example within a system with two coded channels, both channels are active on either different partitions of a core or - given a multi-core system - on multiple cores. 
When one channel fails, the service or application can continue on the other channel. 
The author's approach in \cite{6625234} does improve the MTTF -  which represents the durability of the system - significantly for fail-safe applications.

Another approach presented in \cite{4840571} and later extended by \cite{Bapp} uses a simplex architecture. 
A simplex architecture guarantees safety at an application level by "using simplicity to control complexity" \cite{936213}. 
This architecture uses a safety controller subsystem to ensure stability and a high performance control subsystem. 
The system can then decide to use the safety subsystem in case of a fault or failure while in normal operation it uses the high performance subsystem.  
The authors propose a hardware/software approach to guarantee fail-operational behavior and a fault tolerant system using the simplex architecture. 
The method proposed could handle both logical application level faults and fault dependent layers such as real-time operating systems. 

The simplex architecture was extended in the automotive domain by \cite{OOTB19}. 
In this approach the author uses the simplex architecture presented in \cite{4840571} and  \cite{Bapp} and extends it to the fail-operational context, above the virtualization layer with addition of dynamic reconfiguration. 
The authors show that dynamic reconfiguration of a set functions is possible at a system level and shows that redundancy in safety critical functions with dynamic reconfiguration can reduce the hardware redundancy required in E/E architectures.

\markasnew{Other work} \cite{SRTHG18} \markasnew{addresses the automatic optimization of redundant message routings in automotive ethernet networks to enable fail-operational communication.
By comparison in our work we focus on the mitigation of ECU failures instead of communication component failures.
However, in our approach redundant tasks are distributed over the system with redundant communication routes such that once a task is restarted there is always at least one communication path with preceding and succeeding tasks.
}

Overall, hardware components provided with failure detection and mitigation mechanisms are the base for enabling a dynamic fail-operational solution on software level.
\markasnew{
However, the presented approaches lack flexibility and are not dealing with the problem of providing a dynamic fail-operational behavior for an entire system consisting of many applications distributed over multiple ECUs.
}

\subsection{Graceful Degradation}
\label{sec:graceful}

Another way of solving the fail-operational problem is to apply graceful degradation.
Graceful degradation or functional degradation works by stopping the execution of less important applications in a given system, to keep the more important applications in the system functional \cite{gracefulDegredation}. 
In most cases this means the critical application are kept alive at the expense of non-critical applications.
The advantage over other redundancy approaches such as active redundancy is that hardware costs can be significantly reduced.


This method can be achieved statically or dynamically \cite{7363617}. 
In the static case the system is first analysed and then using scheduling and degradation the fail-operational behaviour is ensured. 
By contrast, in the dynamic case the allocation and reservation of different resources is done at run-time. Therefore, for the dynamic case the system needs to be self-aware at run-time to map its tasks and resources correctly. 

According to \cite{ShKN03} when it comes to graceful degradation, there are two terms that are important to consider: 

\begin{enumerate}
    \item  \textbf{Survivability:} is the dependability factor of how the functionality in the system would be degraded in case of failures \cite{survivability}.
    
    \item \textbf{Performabiltiy:} is the measure of performance and reliability in case of failures \cite{performability}.
    
\end{enumerate}

To ensure fail-operational behavior, the survivability of the system must be ensured even at a lower performability and this can be achieved through graceful degradation. 
In \cite{HW91}, the author formally defines graceful degradation by defining a full list of system constraints that define what tasks the system can perform according to its constraints.
By contrast, in \cite{ShKN03} the authors argues that defining all the constraints in a system is not an easy task and proposes a framework and a utility function that allows for design and analysis of graceful degradation in a decentralized systems. 

\markasnew{The authors in} \cite{BeVo15a} \markasnew{have presented a design-time analysis to find valid application mappings in mixed critical systems. 
Here, applications can have multiple redundancies based on their fail-operational level and the system can be degraded by shutting down optional software components.
By contrast, in our approach, instead of having a limited amount of redundancy, we re-establish lost redundancy after a failover. }

In \cite{641271} and \cite{777447},  the authors focus on degrading the performance of an application in case of a failure instead of shutting down the non-critical applications. 
In addition,  in \cite{Nace} the authors propose a reconfiguration based on the available hardware resources after a failure has occurred. 

In \cite{GLHT09} the authors propose a degradation algorithm based on a Binary Decision Diagram (BDD) data structure and a degradation-aware reliability analysis. 
With multiple degradation modes, there is a constraint set which means that higher modes should have a higher reliability than lower ones. 
A design space exploration is performed to find task activations for every single failure configuration.
The online algorithms then checks if there exists a mapping for the current degradation mode, if not it switches to a lower one and deactivates all functions which are not part of the next degradation mode. 
The disadvantage of this approach is that a task activation has to be calculated for every single failure configuration such that a solution for $2^{|\gls{symb:E}|}$ configurations has to be found with \gls{symb:E} being a set of ECUs.
Furthermore, this solution is not flexible with regards to changes in the software system at run-time. 
When new applications are added all $2^{|\gls{symb:E}|}$ configurations would have to be re-calculated.

\markasnew{Overall, none of the mentioned work related to graceful degradation apart from the work} in\cite{GLHT09} \markasnew{takes reliability into consideration. 
However, analyzing the reliability is an important aspect that needs to be considered when designing a gracefully degrading system such that the effectiveness and side effects of the approach can be taken into account.}

\subsection{Reliability Analysis}\label{sec:reliability}

Many methods for reliable embedded system design have been presented in research. The authors in \cite{birolini2003reliability} give an overview of these methods. 
Traditional reliability analysis methods focus on designing and implementing the system first then doing reliability analysis to meet certain requirements and then redesigning the system if the reliability requirements are not satisfied \cite{364626}.
Reliability can be used as a metric in an optimization problem to find a Pareto-optimized solution along with multiple other design objectives such as area or performance when performing a design space exploration.
Therefore, the system can be optimized before implementation on hardware. 
The authors in \cite{227803} , \cite{1395766} and \cite{1342457} all propose solutions to maximize reliability at design time. 
The author in \cite{1395588} uses reliability as a optimization objective at system level design. 
Here, hardware redundancy is used to deal with faults which in turn increases the costs significantly for ensuring reliability.
The authors in \cite{364626} propose an automatic reliability-aware system synthesis method. 
A multiple objective synthesis approach is presented that considers parameters such as area, costs in addition to reliability which allow for generation of reliable embedded systems.
The authors use a data flow graph and a resource graph to model all alternative architecture implementations. Other research such as \cite{glassBook} which is inspired by \cite{KochBook}, proposes a symbolic reliability analysis of self-healing networks with self-reconfiguration and routing. 
This proposed analytical solution takes the performance and memory constraints into account and illustrates the maximum achievable reliability metrics i.e. MTTF for a given system.

\markasnew{While these approaches present a design time synthesis our agent-based approach finds feasible application mappings at run-time.  
It could be interesting to use reliability as an optimization objective directly in the mapping process in future work as in the presented approaches.
However this would require finding optimized solutions at design time and then finding a feasible mapping at run-time as presented in hybrid mapping approaches such as in} \cite{WWGG18}.
\markasnew{
Instead of optimizing reliability, the focus of our work is the reliability and resource consumption analysis of a graceful degradation approach. 
}


The authors in \cite{GLHT09} use reliability as a metric to optimize an embedded system with multiple degradation modes.
With multiple degradation modes there is a constraint set which means that higher modes should have a higher reliability than lower ones. 
\markasnew{
By comparison our work presents an analysis of a system where mappings are obtained at run-time and applications are evaluated individually instead of grouping them in degradation modes. 
Furthermore, the authors do not present an analysis of the resource consumption of the approach such that no conclusion can be drawn about resource savings compared to an active redundancy approach. 
However, this is a critical aspect that needs to be considered when designing a system such that a proper trade off between resource savings and the impact on reliability can be met.
}

Our base system differs from existing work as we use an agent-based approach which finds applications mappings at run-time.
Here, applications can independent from each other allocate and reserve resources.
Our work aims to examine the impact of graceful degradation on the reliability of a highly distributed system consisting of critical and non-critical applications. 
This has not been considered and analyzed in literature yet.

\section{Application Model and System Architecture}
\label{system_model}

\subsection{System}

\begin{figure}[t]
    \centering
	\includegraphics[width=0.8\linewidth]{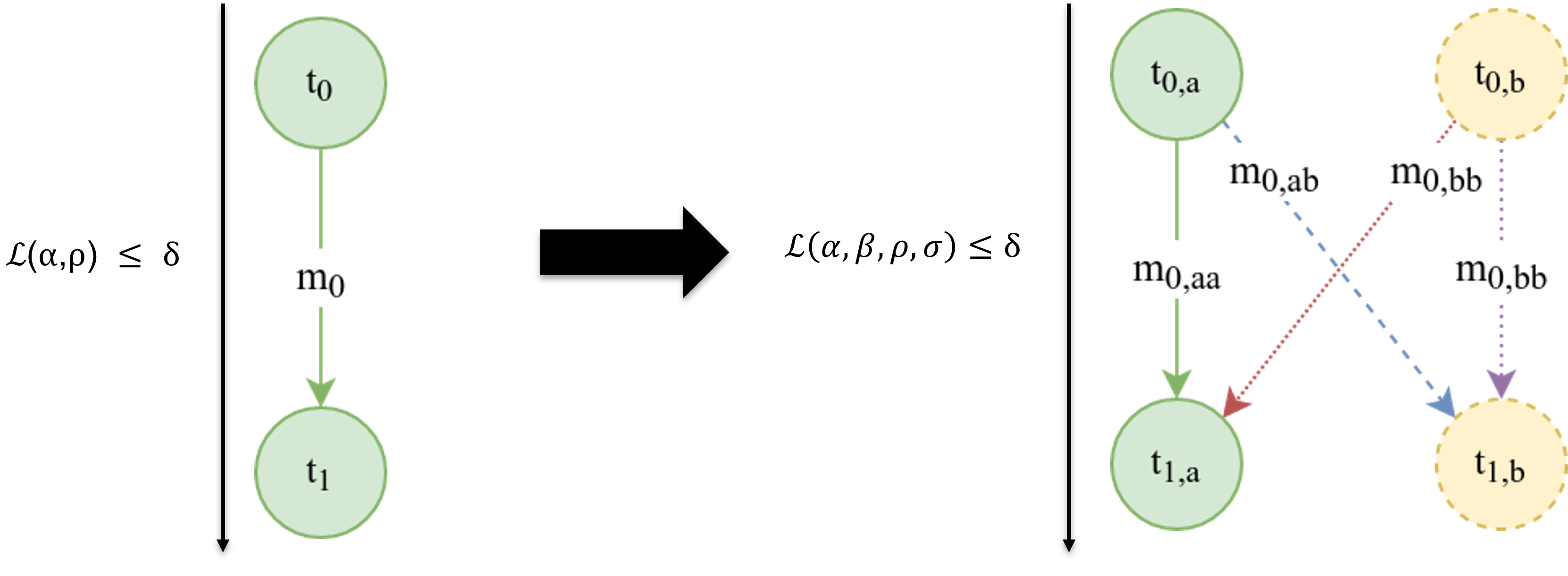}
	\caption{
	\markasnew{An exemplary application graph $G_A(V,E)$ of a non-critical application (left) and a critical application (right).
	The application graph of the critical application consists of two active task instances $t_{0,a}, t_{1,b} \in T_a$ and two passive task instances $t_{0,b}, t_{1,b} \in T_b$. 
	Furthermore, it contains one active message instance $m_{0,aa} \in M_a$ and three backup message instances
	$m_{0,ab}, m_{0,ba}, m_{0,bb} \in M_b$, which are required to ensure there is always a communication path between the tasks instances available.}}
	\label{application_graph}
\end{figure}

\markasnew{In the past, automotive vendors added a new ECU for each new functionality in the vehicle.
Today, cars consist often of more than 100 ECUs to control functions in the domains like infotainment, chassis, powertrain or comfort} \cite{sshzw:2018}.
\markasnew{Now the automotive industry is aiming towards zonal or more centralized architectures.
Some vendors such as Tesla prefer a centralized architecture, where most of the functions are executed on a single ECU such as the FSD computer of Tesla} \cite{Tesla_FSD}.
\markasnew{Bosch is developing a vehicle-centralized, zone-oriented E/E architecture with a few centralized powerful vehicle computers integrating cross-domain functionality similar as} \cite{Bosch_ICAS, Bosch_EE}.
\markasnew{These vehicle computers are connected to actuators and sensors via zone ECUs.
This reduces the required wiring and weight in vehicles but also system complexity.}

\markasnew{In our work we focus on the deployment of bigger applications on a future system architecture which consists of a set of a few ECUs $\gls{symb:e} \in \gls{symb:E}$ which are interconnected via switches and a set of Ethernet links $\gls{symb:l} \in \gls{symb:L}$.
The ECUs and Ethernet links use TDM scheduling with pre-determined time slices which can be allocated or reserved.
To dynamically activate, deactivate, and move tasks on the platform at run-time, we implemented a middleware which is based on SOMEIP} \cite{SOME}, \markasnew{an automotive middleware solution. 
This middleware includes a decentralized service-discovery to dynamically find services in the system and a publish/subscribe scheme to publish and subscribe to events.
}

\subsection{Criticality}
\markasnew{
In our work we are exploring graceful degradation methodologies.
Here, critical applications e.g. for autonomous driving can be restarted after a failure on another ECU. 
Instead of exclusively reserving resources for this scenario, non-critical applications e.g. from the infotainment domain can be shut down to free resources.

According to the ISO 26262 standard, applications can be assigned one of four Automotive Safety Integrity Levels (A to D)} \cite{ISO}.
\markasnew{
However, we do not differ between criticality levels of critical applications in our work as there is no justification in shutting down applications with an assigned ASIL of A for an application with an assigned ASIL of B as the failure of any critical application can have safety-critical consequences.
Instead it has to be ensured that all safety goals are met for any critical application. 
Therefore, we only distinct between critical and non-critical applications.
In our work we assume that each critical application has fail-operational requirements. 
This means that the application has to stay operational even if a failure occurs that affects this application.
}
\begin{itemize}

\item \markasnew{\textit{Critical application}: An application that has fail-operational requirements. 
To ensure a fail-operational behavior passive redundancy on task level is applied. All critical applications have the same priority.}

\item \markasnew{\textit{Non-critical application}: An application without specific safety requirements. 
Non-critical applications can be shut down to free resources for critical applications even if they are not directly affected by a failure.}
\end{itemize}

\subsection{System Software}

Our system software consists of a set of applications $\gls{symb:a} \in \gls{symb:A}$.
We assume that each application $\gls{symb:a}$ is either safety-critical or non-critical.
Applications are executed periodically with a period $\gls{symb:P_a}$ and we assume each application has to meet a deadline $\gls{symb:delta}$, with the period $\gls{symb:P_a}$ being at least as long as the deadline $\gls{symb:delta}$.
For every task we assume that the worst-case execution time (WCET) $\gls{symb:W}(\gls{symb:t})$ is known.
We model each application $\gls{symb:a}$ by an acyclic and directed application instance graph $\gls{symb:G_a}(\gls{symb:vertices},\gls{symb:edges})$, which includes active and passive instances for tasks and messages. 
The vertices $\gls{symb:vertices} = \gls{symb:T_a} \cup \gls{symb:T_p}$ are composed of the set of active task instances $\gls{symb:T_a}$, the set of passive task instances $\gls{symb:T_p}$, the edges $\gls{symb:edges} = \gls{symb:M_a} \cup \gls{symb:M_b}$ are composed of the set of active message instances $\gls{symb:M_a}$ and the set of backup message instances $\gls{symb:M_b}$. 

A binding $\gls{symb:alpha}: \gls{symb:T_a} \rightarrow \gls{symb:E}$  assigns an active task instance $\gls{symb:t} \in \gls{symb:T_a}$ to an ECU $\gls{symb:alpha}(\gls{symb:t}) \in \gls{symb:E}$. Safety-critical application have to fulfill fail-operational requirements and, thus, have to remain operational even during critical ECU failures.
Therefore, we assume that a redundant passive task instance is required for our safety-critical applications.
Here, the binding $\gls{symb:beta}: \gls{symb:T_p} \rightarrow \gls{symb:E}$ assigns a passive task instance $\gls{symb:t} \in \gls{symb:T_p}$ to an ECU $\gls{symb:beta}(\gls{symb:t}) \in \gls{symb:E}$. 
A routing $\gls{symb:rho}: \gls{symb:M_a} \rightarrow 2^{\gls{symb:L}}$ assigns each active message $\gls{symb:m} \in \gls{symb:M_a}$ to a set of connected links $\gls{symb:L}' \subseteq \gls{symb:L}$ that establish a route $\gls{symb:rho}(\gls{symb:m})$. 
We use the shortest path routing obtained through Dijkstra's algorithm such that there is only a single route between two ECUs $e$ available \cite{Dijkstra}.
Similar to the binding, up to three backup routes are required of which one will get activated after an ECU failure depending on which passive task instances get activated.
Here, a routing $\gls{symb:sigma}: \gls{symb:M_b} \rightarrow 2^{\gls{symb:L}}$ is assigned to each backup message $\gls{symb:m} \in \gls{symb:M_b}$.

In this work and our experiments we focus on mitigating ECU failures which can be detected by watchdogs and heartbeats. 
However, our approach is also able to mitigate single task failures when using additional failure detection mechanisms.
Our graceful degradation approach ensures that safety-critical applications can keep running after an ECU failure while there is no guarantee for non-critical applications.
Non-critical task might even be shut down to free resources to free sufficient resources to start critical passive task instances.
In the case of an ECU failure we consider that the current application execution might not finish if an active task instance is affected directly by the failure and that application execution might be interrupted for a certain time interval \cite{WEWS21}.
Checkpoints can be periodically transmitted from active to passive task instances to save important state data \cite{weiss2021checkpointing}.
After the failure recovery computation can be continued with the latest transmitted checkpoints. 

Figure \ref{application_graph} \markasnew{presents a non-critical and a critical application according to our system model.
The non-critical application consists of two tasks and one message being sent between the two tasks. 
For the safety-critical application the graph also includes two passive task instances and three backup message instances. 
Three backup message instances are required such that it can be ensured that always a communication between two task instances is possible regardless of which task instances are affected by a failure. 
The message instances $m_{0,ab}$ and $m_{0,ba}$ are required if only one of the active task instances is failing, while the message instance $m_{0,bb}$ is required if both active task instances are affected by a failure, e.g. because they are mapped onto the same failing ECU.}

\begin{figure}[t]
    \centering
	\includegraphics[width=0.8\linewidth]{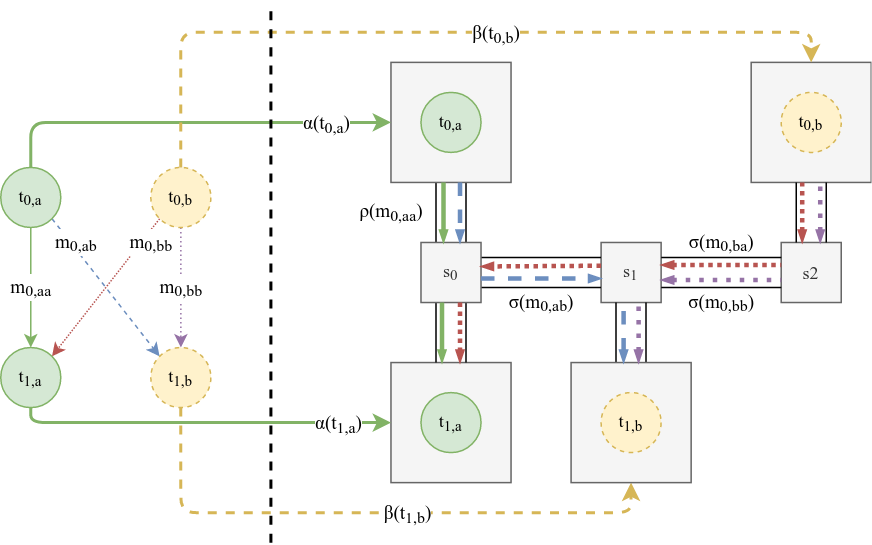}
	\caption{
	\markasnew{Exemplary mapping of a safety-critical application onto a hardware architecture consisting of four ECUs $e_0$, $e_1$, $e_2$, and $e_3$, and three switches $s_0$, $s_1$, and $s_2$.
	The green arrows indicate the active bindings of tasks $t_0$ and $t_1$, while the dashed yellow arrows indicate the passive task bindings.
	The routings of the message instances are indicated by the same arrow color and style as in the application graph.}}
	\label{fig_mapping}
\end{figure}

Figure \ref{fig_mapping} \markasnew{shows the binding $\alpha$ of the active task instances $t \in T_a$ and the binding $\beta$ of the passive task instances $t \in T_b$ onto a system architecture. 
The routing $\rho$ of the active message instance $m \in M_a$ and the routings $\sigma$ of the backup message instances $m \in M_b$ are also marked by colored arrows.}

\section{Gracefully Degrading System Architecture}
\label{graceful}
\markasnew{In this section we introduce our existing graceful degradation approach from} \cite{weiss2023predictable} \markasnew{which is based on composable scheduling where applications are isolated such that they do not influence each other allowing to verify their timing behavior independently.}
However, in the work at hand we are not focusing on timing analysis but make use of our existing approach on which we perform our reliability analysis and which we use for our experiments.

\subsection{Composable Scheduling of Gracefully Degrading Systems}
We extend state-of-the-art TDM scheduling by introducing the concept of graceful degradation to CPU resources.
Applied to our composable schedule, this means that slots can not only be allocated for a task but also reserved.
A reservation indicates that the corresponding slot is currently not in use, but might be used and turned into an allocation once the passive task instance is being used.
slots that can be reserved are empty slots that have not been allocated yet or slots which are already allocated by non-critical applications.
The allocation of service-intervals works vice-versa, non-critical applications can allocate slots that are free or which are already reserved by critical applications.
Critical applications on the other hand can only allocate slots that are completely free.
If a slot is allocated by a non-critical application and also reserved by a critical application the graceful degradation approach is applied.
In case there is a failure in the system and critical passive task instances have to be started to mitigate a failure, the reservation of the resources will be turned into an active allocation and any non-critical tasks which formerly hold an allocation of the corresponding slots are degraded.
Depending on the application, it could then be decided if the degraded non-critical application keeps running in a degraded mode or is completely shut down.

\begin{figure*}[ht]
\subfloat[Task schedule with reservation and allocation.]{
	\includegraphics[width=0.3\linewidth]{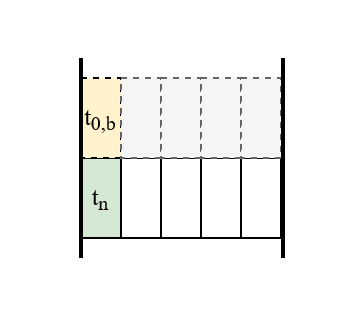}
}
\hfill
\subfloat[Non-critical task $t_n$ being shut down.]{
	\includegraphics[width=0.3\linewidth]{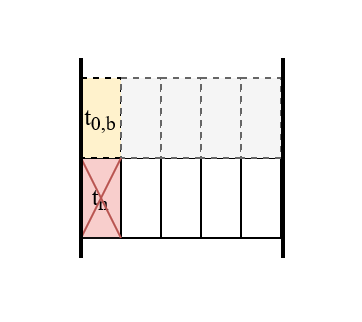}
}
\hfill
\subfloat[Reservation turned into an allocation.]{
	\includegraphics[width=0.3\linewidth]{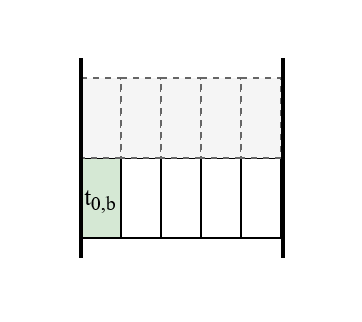}
}
\caption{Example of a task schedule with a maximum amount of allocatable (lower) and reservable (upper) slots of $\gls{symb:SI_{max}} = 5$. 
One slot is reserved for the critical task instance $\gls{symb:t}_{0,b}$. 
The same slot is allocated by the non-critical active task instance $t_n$.
In a failure scenario where $\gls{symb:t}_{0,b}$ has to be activated, $t_n$ is being shut down in a first step.
Afterwards, $\gls{symb:t}_{0,b}$ is taking over the allocation of the slot.
}
\label{task_schedule_degradation}
\end{figure*}

In Figure \ref{task_schedule_degradation} an exemplary task schedule with reservations and allocations is presented. The upper slots indicate a reservation, while the lower slots indicate an allocation of the same slot. 
In this example the first slot is allocated by the non-critical task $t_n$ but also reserved by the critical task instance $\gls{symb:t}_{0,b}$.
In a failure scenario where the task instance $\gls{symb:t}_{0,b}$ is activated to serve as a backup solution, the non-critical $t_n$ first looses its allocation and, thus, is being degraded.
Afterwards, the reservation of the task instance $\gls{symb:t}_{0,b}$ is turned into an active allocation such that this resource can now be used exclusively by the critical task instance.

\subsection{Strategies}
\label{section_strategies}
\markasnew{The decision which slot should be allocated or reserved is not straightforward.
When always a completely free slot is taken first, then the schedule might run out of slots earlier such that not sufficient resources might be left for other tasks. 
On the other hand, when it is tried to overlap slots as much as possible but more than sufficient slots are available for all tasks, then an avoidable degradation might occur in a failure scenario.
In the following we discuss the three strategies we developed to allocate and reserve resource slots and evaluate them in Section} \ref{evaluation}.

Changing this algorithm can impact the degradation behaviour and success chance of finding a mapping.
In the following we propose introduce the three strategies \emph{Random}, \emph{FreeFirst}, and \emph{FreeLast} for assigning slots.

\markasnew{The \emph{FreeFirst} strategy aims at minimizing the overlap between reservations and allocations by assigning slots first that have not been allocated or reserved. 
Only if no free slot is available the algorithm will allocate or reserve other slots.
The algorithm has the advantage of reducing the degradation effect as allocations and reservations overlap as little as possible.
The downside is that in scenarios where resources are constrained, the algorithm might lead to lower success rates of finding mappings as less free slots will be available. 
On the other hand, in more relaxed scenarios it uses all available resources to reduce the degradation effect.}

\markasnew{The \emph{FreeLast} strategy aims at utilizing resources more efficiently by assigning slots first that already have been allocated or reserved, which maximizes the overlap between reservations and allocations.
The advantage is that less slots are required compared to the other two strategies.
On the other hand, as the overlap between reservations and allocations is maximized, there will be a stronger degradation effect in case of a failure scenario.
In scenarios where resources are less constrained, reservations and allocations will overlap even if additional free slots were available leading to avoidable degradation effects.}

\markasnew{When choosing one of the two opposing strategies, a trade-off between degradation effect, success rate and resource-efficiency has to be made.
In scenarios where degradation is desired or tolerated, the \emph{FreeLast} strategy can lead to a lower resource utilization.
In scenarios where many resources are available and degradation should be avoided as much as possible the \emph{FreeFirst} strategy maximizes resource utilization to minimize degradation impact.
We continue this discussion with our experimental results in Section} \ref{evaluation}, \markasnew{which gives further insights into these three strategies.
}

\section{Reliability Analysis}
\label{reliability_analysis}


In this section we are giving a short introduction into reliability analysis. 
Afterwards, we present our approach to formally analyze the impact of graceful degradation on the reliability of critical and non-critical applications.
We use these structure functions to evaluate our graceful degradation approach in Section \ref{evaluation}.

\subsection{Introduction}

Given a specified amount of time, the reliability $\gls{symb:R}(\gls{symb:tau})$ of a system is the probability that the system can operate continuously without failures.
Usually the reliability of the system is defined  in a time interval of zero to \gls{symb:tau} where \gls{symb:tau} can be any specified amount of time in the future up to infinity. The reliability of the system is complimentary to its failure rate denoted as $\gls{symb:lambda}(\gls{symb:tau})$:

\begin{equation}
\label{eq:failure}
\gls{symb:lambda}(\gls{symb:tau}) = 1-\gls{symb:R}(\gls{symb:tau})
\end{equation}

Reliability is usually expressed as a distribution function. Electronic components and systems usually have an exponential distribution function as the failure rate $\gls{symb:lambda}(\gls{symb:tau})$. The exponential distribution is the only distribution to have a constant failure rate. This simplifies things with having only one unknown variable which is the failure rate. 
Therefore, the reliability of an exponential system can the be expressed as:
\begin{equation}
\label{eq:ReliabilityEquation}
  \gls{symb:R}(\gls{symb:tau}) = e^{-\int_{0}^{\gls{symb:tau}}\gls{symb:lambda}(\theta)d\theta}
\end{equation}

The failure rate $\gls{symb:lambda}(\gls{symb:tau})$ of a large enough sample of independent components can be represented with the bathtub curve \cite{birolini2003reliability}. 
The bathtub curve defines the failure rate of the exponential distribution in three main phases:
\begin{itemize}
    \item Phase one refers to initial manufacturing defects at time \gls{symb:tau}=0 with a high failure rate due to random weaknesses such as production defects or design errors. During this period the failure rate drops rapidly with time. This stage usually is referred to as infant mortality.
    \item Phase two is the operational phase and usually has an approximate constant failure rate. The failures here happen due to external factors such as collisions, overloading or human error. This stage is referred to as the useful life of the component.
    \item Phase three is the final phase and will see a significant increase in failure rate over time. This is due to ageing effects of the electronic components and other factors that lead to a failure. This stage is referred to as wear-out period. 
\end{itemize}

In most cases when performing reliability analysis it is assumed that the system is faultless at start time which means $\gls{symb:lambda}(0) = 1$. This further simplifies the reliability equation to

\begin{equation}
\label{eq:SimplifiedReliabilityEquation}
 \gls{symb:R}(\gls{symb:tau})	=e^{-\gls{symb:lambda} \gls{symb:tau}}
\end{equation}

For a given time $\gls{symb:tau}$ the only unknown factor is the failure rate  $\gls{symb:lambda}$.
In our case the failure rate of each component (ECU) is a constant arbitrary value as all of our components are abstracted and are assumed to have an identical failure rate.
In our experiments we set the component failure rate to $\gls{symb:lambda} = 0.01$. 

Equation \ref{eq:SimplifiedReliabilityEquation} represents the reliability distribution function of one ECU in our system. 
We use this reliability function to derive the Mean-Time-To-Failure (MTTF) as a metric for the reliability of both critical and non-critical applications.
The MTTF represents the average lifespan of a component or system from the moment it was installed until a failure occurs. 
Therefore, given the calculations and assumptions stated previously the MTTF is calculated as:
\begin{equation}
\label{eq:MTTF}
\gls{symb:MTTF}(\gls{symb:tau}) = e^{-\int_{0}^{\infty}\gls{symb:lambda}(\theta)d\theta} =\frac{1}{\gls{symb:lambda} (\gls{symb:tau})}.
\end{equation}

\subsection {Structure Functions}
\markasnew{To describe the behavior of our applications we use boolean functions represented by a structure function $\gls{symb:varphi}$, which is encoded in a Binary Decision Diagram (BDD)} \cite{364626}.
\markasnew{A BDD is a rooted, directed and acyclic graph which consists of multiple decision nodes and two terminal nodes which determine the outcome of the boolean function} \cite{JREL08}.
\markasnew{The maximal size of a BDD encoding a formula on $n$ variables is $2^{n}/n$} \cite{rauzy1993new}. 
\markasnew{The complexity of automatically generating our structure functions as described below is linear on the number of tasks.}

\marknew{
In a system with $n$ components the state of each component $i$ can be encoded in a boolean variable $x_i$, which evaluates to 1 if the component is operational and to 0 if the component is defect.
A structure function $\gls{symb:varphi}(\gls{symb:x}) = \gls{symb:varphi}(\gls{symb:x}_1, \gls{symb:x}_2, ..., \gls{symb:x}_n)$ which represents the state of the system evaluates to 1 if the system is operational and to 0 if not. 

In the following we denote $\gls{symb:z}: \gls{symb:vertices} -> \{0,1\}$ as a function, which translates the task instances into a binary variable with 1 indicating a proper operation of the task instance.
We define that $\gls{symb:varphi}(\gls{symb:z})=\gls{symb:varphi}(\gls{symb:z}(\gls{symb:t}_{0,a}),\gls{symb:z}(\gls{symb:t}_{0,b}), ... \gls{symb:z}(\gls{symb:t}_{n,a}),\gls{symb:z}(\gls{symb:t}_{n,b}))$ represents the structure function of an application $a$.
To derive the reliability, we translate these boolean variables which are dependent on the state of the task instances into boolean variables which are only dependent on the state of the ECUs.
Therefore, we denote $\gls{symb:y}: \gls{symb:E} -> \{0,1\}$ as a function which translates ECUs into a binary variable with 1 indicating a proper operation of the ECU.
Furthermore, we define $\gls{symb:u}: \gls{symb:vertices} \rightarrow E$ as a function which translates a boolean variable $\gls{symb:z}(\gls{symb:t})$ dependent on the state of a task instance $\gls{symb:t}$ to a boolean variable $\gls{symb:y}(\gls{symb:e})$ which only depends on the state of ECU $\gls{symb:e}$, such that
$\gls{symb:u}(\gls{symb:z}(\gls{symb:t})) = \gls{symb:y}(\gls{symb:e}) $, where $\gls{symb:alpha}(\gls{symb:t}) = \gls{symb:e}$ for active task instances and $\gls{symb:beta}(\gls{symb:t}) = \gls{symb:e}$ for passive task instances.
As an example, if a task instance $\gls{symb:t}_{0,a}$ is bound to an ECU $\gls{symb:alpha}(\gls{symb:t}_{0,a})=\gls{symb:e}_1$, the function $\gls{symb:z}(\gls{symb:t}_{0,a})$ only evaluates to 1 as long as the function $\gls{symb:y}(\gls{symb:e}_1)$ is indicating a proper operation.
The resulting structure function $\gls{symb:varphi}(\gls{symb:u}(\gls{symb:z}))=\gls{symb:varphi}(\gls{symb:y})=\gls{symb:varphi}(\gls{symb:y}(\gls{symb:e}_0), ... \gls{symb:y}(\gls{symb:e}_{n}))$ is then only dependent on a set of independent variables indicating the status of the ECUs.
}


\subsection{Derivation of Reliability}
\markasnew{In our experiments we use the JRELIABILITY framework to evaluate the MTTF} \cite{JREL08}. 
\markasnew{Here, a Shannon-decomposition based algorithm is applied to the structure function to calculate the reliability} \cite{364626}.
\markasnew{The complexity of this algorithm is linear in the size of the BDD} \cite{rauzy1993new}.
\marknew{
A BDD allows to calculate the probability of the root event of the the tree using Shannon's decomposition if the probabilities of the leaves are given} \cite{rauzy1993new} \marknew{as follows:}

\begin{equation}
\gls{symb:p}(\gls{symb:f}) = \gls{symb:p}(\gls{symb:x} = 1) \cdot \gls{symb:p}(\gls{symb:f}_{\gls{symb:x}=1}) + \gls{symb:p}(\gls{symb:x} = 0) \cdot \gls{symb:p}(\gls{symb:f}_{\gls{symb:x}=0}),
\end{equation}

\marknew{
with $\gls{symb:p}$ being the probability that is calculated, $\gls{symb:f}$ being a function and $\gls{symb:x}$ a variable occurring in $\gls{symb:f}$.

Using the probability function $\gls{symb:P}(\gls{symb:varphi}, \gls{symb:tau})$ as the probability $p$ being calculated and the structure function $\gls{symb:varphi}$ as the function $f$ being evaluated, the probability of a working system can be calculated as} \cite{glass2010towards}:
\begin{equation}
\gls{symb:P}(\gls{symb:varphi}, \gls{symb:tau}) = \gls{symb:P}(\gls{symb:y}(\gls{symb:e})=1, \gls{symb:tau}) \cdot \gls{symb:P}(\gls{symb:varphi}_{\gls{symb:y}(\gls{symb:e})=1}, \gls{symb:tau}) + \gls{symb:P}(\gls{symb:y}(\gls{symb:e})=0, \gls{symb:tau}) \cdot \gls{symb:P}(\gls{symb:varphi}_{\gls{symb:y}(\gls{symb:e})=0}, \gls{symb:tau}).
\end{equation}
\marknew{
With a homogeneous set of ECUs and the assumption that the same reliability function $\gls{symb:R}(\gls{symb:varphi})$ is applied to each ECU, this can be further simplified to:
}
\begin{equation}
\gls{symb:P}(\gls{symb:varphi}, \gls{symb:tau}) = \gls{symb:R}( \gls{symb:tau}) \cdot \gls{symb:P}(\gls{symb:varphi}_{\gls{symb:y}(\gls{symb:e})=1}, \gls{symb:tau}) + (1 - \gls{symb:R}(\gls{symb:tau})) \cdot \gls{symb:P}(\gls{symb:varphi}_{\gls{symb:y}(\gls{symb:e})=0}, \gls{symb:tau}),
\end{equation}
\marknew{
resulting in the desired reliability function $\gls{symb:R}(\gls{symb:varphi}, \gls{symb:tau}) = \gls{symb:P}(\gls{symb:varphi}, \gls{symb:tau})$ describing the probability of a proper working application.
Finally, using equations}\ref{eq:SimplifiedReliabilityEquation} and \ref{eq:MTTF}, \marknew{the MTTF can be derived from $\gls{symb:R}(\gls{symb:varphi}, \gls{symb:tau})$.}


\subsection{Structure Functions - Graceful Degradation Approach}

In the following we assume that a mapping process including the allocation and reservation has been performed.
The mapping of the tasks as well as the information which specific slots have been allocated and reserved for each task are required to determine the structure function $\gls{symb:varphi}$ for each application.
While the derivation of the structure function $\gls{symb:varphi}$ is more straight-forward for critical applications, it is more complex for non-critical applications as the reliability is also influenced by the degradation process.
We assume that an application is still operational as long as there is at least one task instance of each task available and still able to communicate.

\subsubsection{Structure Function of Critical Applications}

\begin{figure*}[ht]
\subfloat[Mapping example. \label{bdd_critical_mapping}]{
	\includegraphics[width=0.75\linewidth]{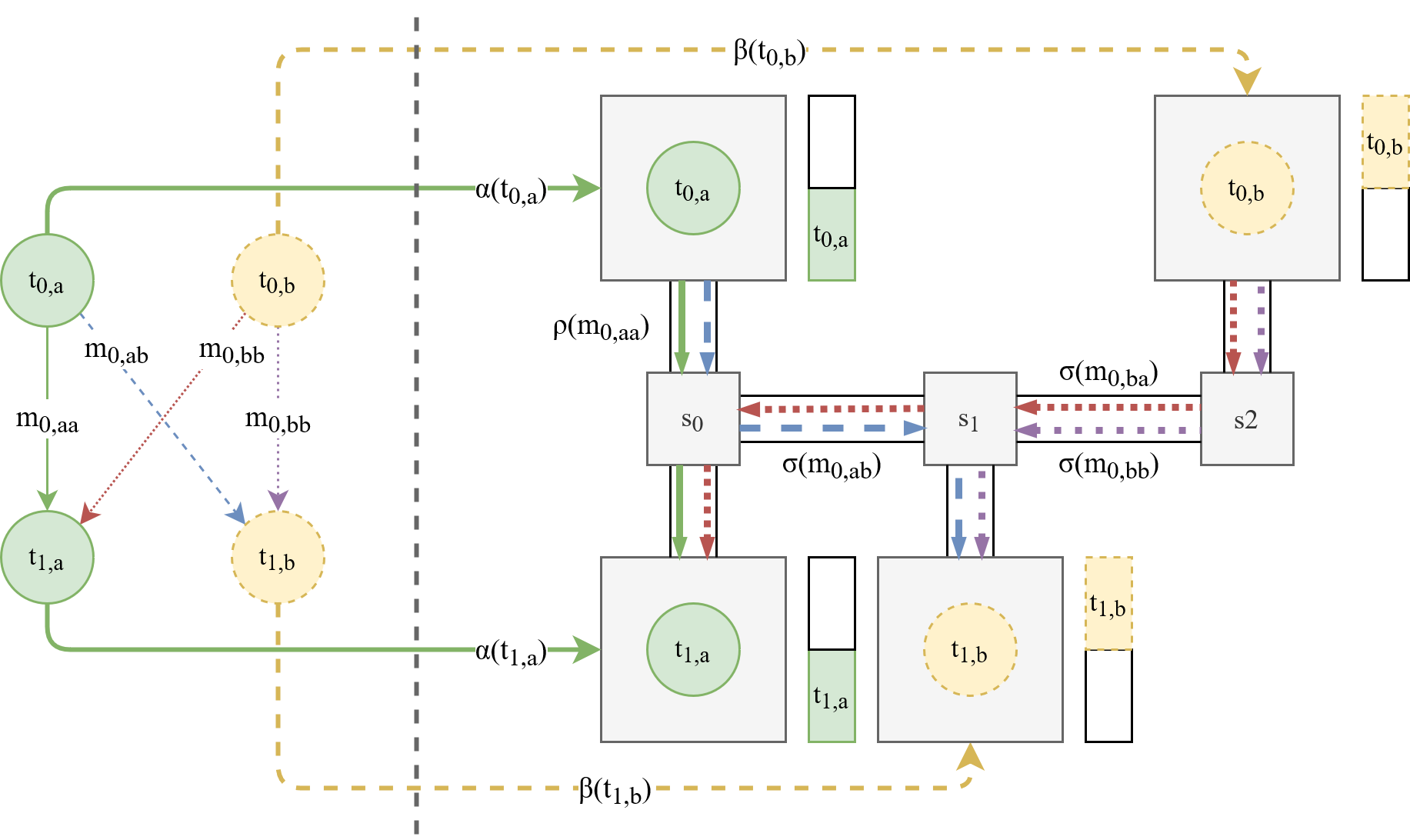}
}
\hfill
\subfloat[BDD.  \label{bdd_critical}]{
	\includegraphics[width=0.19\linewidth]{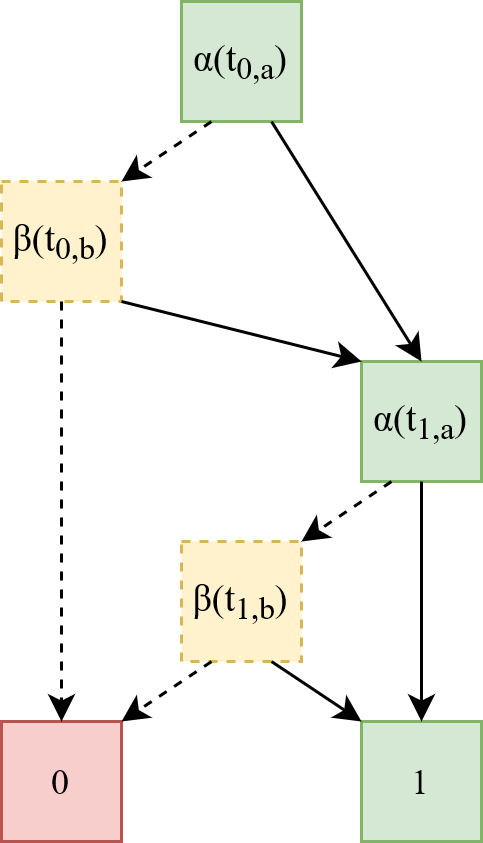}
}
\caption{
An example mapping of a critical application onto four ECUs and the corresponding BDD. The slot allocation and reservation is indicated right of the ECUs. The BDD represents the boolean structure function $\gls{symb:varphi}(\gls{symb:y}) = (\gls{symb:y}(\gls{symb:alpha}(\gls{symb:t}_{0,a})) \vee \gls{symb:y}(\gls{symb:beta}(\gls{symb:t}_{0,b}))) \wedge (\gls{symb:y}(\gls{symb:alpha}(\gls{symb:t}_{0,a})) \vee \gls{symb:y}(\gls{symb:beta}(\gls{symb:t}_{1,b})))$. If both task instances of any task where affected by ECU failures, then the application would fail.
}

\end{figure*}

Critical applications in our system model are required to have double redundancy which means every active task instance has a passive counterpart. 
As the active task instance and passive tasks instances are mapped onto different ECUs, a critical application would always stay operational in the event of a single ECU failure.
However, if two or more ECUs had a critical failure and both task instances of a task are affected by it, then the critical application fails.

As an example a critical application consisting of two active and two passive task instances is mapped on to four ECUs in Subfigure \ref{bdd_critical_mapping}.
One slot has been allocated or reserved for the active and passive task instances.
If both the active and passive task instances $\gls{symb:t}_{0,a}$ and $\gls{symb:t}_{0,b}$ or $\gls{symb:t}_{1,a}$ and $\gls{symb:t}_{1,b}$ are affected by ECU failures then the application would be considered as failed, resulting in the structure function:

\begin{equation}
    \gls{symb:varphi}(\gls{symb:z}) = (\gls{symb:z}(\gls{symb:t}_{0,a}) \vee \gls{symb:z}(\gls{symb:t}_{0,b})) \wedge (\gls{symb:z}(\gls{symb:t}_{0,a}) \vee \gls{symb:z}(\gls{symb:t}_{1,b})).
\end{equation}

The resulting structure function $\gls{symb:varphi}(y)$ dependent on the operational status of the ECUs is:
\begin{equation}
    \gls{symb:varphi}(\gls{symb:y}) = (\gls{symb:y}(\gls{symb:alpha}(\gls{symb:t}_{0,a})) \vee \gls{symb:y}(\gls{symb:beta}(\gls{symb:t}_{0,b}))) \wedge (\gls{symb:y}(\gls{symb:alpha}(\gls{symb:t}_{0,a})) \vee \gls{symb:y}(\gls{symb:beta}(\gls{symb:t}_{1,b})))
\end{equation}

Subfigure \ref{bdd_critical} shows the corresponding BDD, with all possible paths leading either to a failure (0) or to success (1). 
Each decision node in the BDD has two outgoing edges that correspond to the variable being 0 (dashed arrow) or 1 (normal arrow).
Each variable assignment that results in 1 means that the application is still operational, while the paths leading to 0 represent a failed application.

Generalizing the structure function $\gls{symb:varphi}(z)$ for critical applications, if for all tasks $\gls{symb:t} \in \gls{symb:vertices}$  either the active task instance $\gls{symb:t}_{a}$ or the the passive task instance $\gls{symb:t}_{b}$ is working, then the critical application is operational:

\begin{equation}
\label{structure_function_critical}
    \gls{symb:varphi}(\gls{symb:z}) = \bigwedge_{\gls{symb:t} \in \gls{symb:vertices}} \gls{symb:z}(\gls{symb:t}_{a}) \vee \gls{symb:z}(\gls{symb:t}_{b}).
\end{equation}

\subsubsection{Structure Function of Non-Critical Applications}

\begin{figure*}[ht]
\subfloat[Mapping example. \label{bdd_non_critical_mapping}]{
    
	\includegraphics[width=0.73\linewidth]{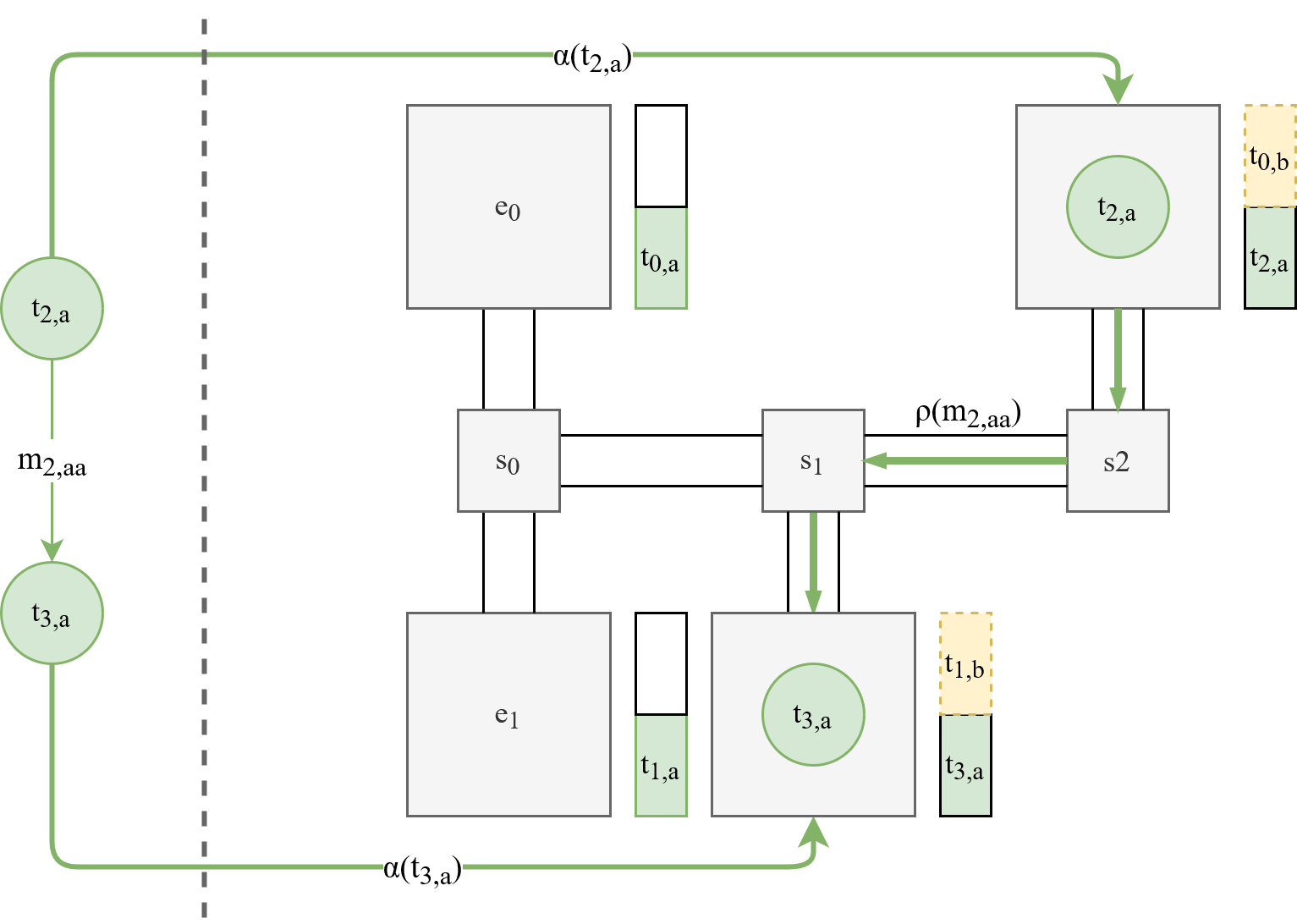}
}
\hfill
\subfloat[BDD. \label{bdd_non_critical}]{
	\includegraphics[width=0.21\linewidth]{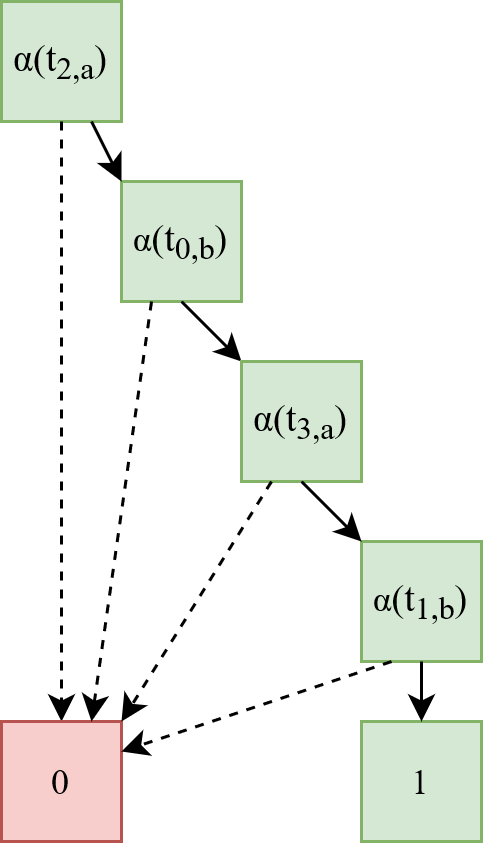}
}
\caption{
An example mapping of a non-critical application onto the system and the corresponding BDD. 
The slot allocation and reservation is indicated right of the ECUs. 
The slots which are allocated by the tasks instances are also reserved by the passive task instances of the critical example from the previous example.
The BDD represents the boolean structure function $\gls{symb:y}(\gls{symb:alpha}(\gls{symb:t}_{2,a})) \wedge \gls{symb:y}(\gls{symb:alpha}(\gls{symb:t}_{0,a})) \wedge \gls{symb:y}(\gls{symb:alpha}(\gls{symb:t}_{3,a})) \wedge \gls{symb:y}(\gls{symb:alpha}(\gls{symb:t}_{1,a}))$. 
If any of the task instances were affected by an ECU failure the non-critical application would immediately fail. 
Furthermore, a degradation effect has to be considered as the the slots of the task instances are reserved by passive passive task instances of the critical application,
If any of the the active task instances of the critical applications were affected by a failure, the corresponding passive task instance would get activated and the task instance of the non-critical application would be degraded.
}
\end{figure*}

As we assume that no redundancy is used, a non-critical applications is not operational anymore if any of the tasks is affected by a failure. 
However, not only a direct ECU failure can lead to an application failure but also degradation effects. 
If a slot which is allocated by a non-critical task is reserved by passive task instance of a critical application then a degradation occurs if that passive task instances is activated. 
An activation of this passive task instance would only occur if the corresponding active task instance was affected by an ECU failure.
This means that the failure of an ECU to which none of the tasks of a non-critical application are mapped to, can indirectly lead to a failure of that application through degradation.

As an example Subfigure \ref{bdd_non_critical_mapping} presents the mapping of a non-critical application consisting of the task instances $\gls{symb:t}_{2,a}$ and $\gls{symb:t}_{3,a}$ onto the system.
In this example we assume that the critical application from the previous example in Subfigure \ref{bdd_critical_mapping} is also still mapped onto ECUs.
The two task instances allocated a slot each.
However, both of these slots are also reserved by the passive task instances $\gls{symb:t}_{0,b}$ and $\gls{symb:t}_{1,b}$ of the critical application. 
Here, if either the ECU of the task instance $\gls{symb:t}_{0,a}$ or the ECU of the task instance $\gls{symb:t}_{1,a}$ of the critical application failed, then the corresponding passive task instance $\gls{symb:t}_{0,b}$ or $\gls{symb:t}_{1,b}$ would be activated.
This would then lead to a shutdown of either $\gls{symb:t}_{2,a}$ or $\gls{symb:t}_{3,a}$ and, therefore, to an indirect shutdown of the non-critical application.
The structure function \gls{symb:varphi}(\gls{symb:z}) is:
\begin{equation}
    \gls{symb:varphi}(\gls{symb:z}) = \gls{symb:z}(\gls{symb:t}_{2,a}) \wedge \gls{symb:z}(\gls{symb:t}_{0,a}) \wedge \gls{symb:z}(\gls{symb:t}_{3,a}) \wedge \gls{symb:z}(\gls{symb:t}_{1,a})
\end{equation}

The resulting structure function \gls{symb:varphi}(\gls{symb:y}) is is also represented as a BDD in Subfigure \ref{bdd_non_critical}:
\begin{equation}
    \gls{symb:varphi}(\gls{symb:y}) = \gls{symb:y}(\gls{symb:alpha}(\gls{symb:t}_{2,a})) \wedge \gls{symb:y}(\gls{symb:alpha}(\gls{symb:t}_{0,a})) \wedge \gls{symb:y}(\gls{symb:alpha}(\gls{symb:t}_{3,a})) \wedge \gls{symb:y}(\gls{symb:alpha}(\gls{symb:t}_{1,a})) = \gls{symb:y}(\gls{symb:e}_2) \wedge \gls{symb:y}(\gls{symb:e}_1) \wedge \gls{symb:y}(\gls{symb:e}_3) \wedge \gls{symb:y}(\gls{symb:e}_0)
\end{equation}

Overall, this means that the failure of any ECU in this example would lead to a failure of the non-critical application. 
In exchange the critical application is still able to stay operational if only one ECU is affected by a failure.
Therefore, through  passive redundancy and graceful degradation, the reliability of the critical application is increased at the cost of reducing the reliability of the non-critical application.

The generalized structure function $\gls{symb:varphi}(\gls{symb:z})$ for non-critical applications is:

\begin{equation}
\label{structure_function_non_critical}
    \gls{symb:varphi}(\gls{symb:z}) = \bigwedge_{\gls{symb:t} \in \gls{symb:vertices}}(\gls{symb:z}(\gls{symb:t}_{a}) \wedge \bigwedge_{\gls{symb:t}^r \in \gls{symb:specialset}}\gls{symb:z}(\gls{symb:t}_{a}^r)).
\end{equation}

The first part of the equation corresponds to the direct failure of the ECU $\gls{symb:alpha}(\gls{symb:t}_{a})$ to which the the task $\gls{symb:t}_{a}$ is mapped to.
The second part of the equation includes the indirect shutdown through degradation.
Here we define that that $\gls{symb:t}^r \in \gls{symb:specialset}$ includes all tasks $\gls{symb:t}^r$ of critical applications where a passive task instance has reserved a slot that is also allocated by the task $\gls{symb:t}_{a}$.
The failure of any ECU $\gls{symb:alpha}(\gls{symb:t}_{a}^r)$ would lead to an activation of the corresponding passive task instance $\gls{symb:t}_{b}^r$ and to a shutdown of the task $\gls{symb:t}_{a}$.

Using the reliablity function \ref{eq:SimplifiedReliabilityEquation} and the structure functions \ref{structure_function_critical} and \ref{structure_function_non_critical}, the reliability function and MTTF of each critical and non-critical application can be finally obtained. We are using these functions and the MTTF to evaluate the impact of graceful degradation on non-critical applications in Section \ref{evaluation}.

\section{Evaluation}
\label{evaluation}

In the following we analyze the impact of graceful degradation on the reliability of critical and non-critical applications using our in-house developed simulation framework. 
We first introduce our experimental setup in Subsection \ref{setup} .
Afterwards we present experimental results of our graceful degradation approach and compare it to active redundancy in Subsection \ref{experiment_default}.
Then we analyze the effect of our allocation and reservation strategies on the reliability in Subsection \ref{experiments_strategies}.
In Subsection \ref{experiments_exposure} we present and analyze results obtained with the \textit{Predecessor Heuristic}, which aims at reducing the failure exposure of the applications.
Last, we summarize all findings from these experiments and draw a conclusion about the importance and impact of graceful degradation in Subsection \ref{summary_experiments}.

\subsection{Experimental Setup}
\label{setup}
Our simulation framework has been developed to simulate automotive hardware architectures and the execution and communication of the system software according to our system model in Section \ref{system_model}.
On top of the simulation framework we implemented the agents, resource managers and strategies.
For the simulation framework we chose a process-based Discrete-Event Simulation (DES) architecture based on the SimPy framework \cite{Simpy}.
The hardware architecture and system software are described in a specification file using the XML schema from the OpenDSE framework \cite{RLGS19}.
The simulation framework supports any kind of hardware architecture consisting of ECUs, switches, and links.
To allow a dynamic behavior where tasks and agents are moving between ECUs at run-time we use a communication middleware based on the SOME/IP standard \cite{SOME}.
The middleware consists of a service discovery which allows to dynamically find services at run-time.
Communication participants are either modelled as clients or services. 
Furthermore, the middleware supports remote-procedure calls and includes a publish/subscribe scheme.
The framework also offers the possibility to simulate ECU failures by shutting down ECUs. 
ECU failures are detected via heartbeats that are periodically sent between all ECUs.
Once a watchdog does not receive the heartbeat within a certain timeout interval it reports the corresponding ECU failure.
We use shortest path routing based on Dijkstra's algorithm for routing the messages \cite{Dijkstra}.
For our experiments we use applications which are synthetically generated by the OpenDSE framework using the TGFF algorithm \cite{RLGS19, Dick98}. 

After each application has been mapped successfully, our framework generates the structure functions according to equations \ref{structure_function_critical} and \ref{structure_function_non_critical}. 
Afterwards, we use the JRELIABILITY framework \cite{JREL08} to evaluate the structure functions and to obtain the MTTF as our evaluation metric.
In the following we calculate and present the average MTTF of the critical and non-critical applications as

\begin{equation}
\label{eq:AVG_MTTF}
\gls{symb:MTTF_{AVG}} = \frac{\sum_{i=1}^{n} \gls{symb:MTTF}_i}{n},
\end{equation} 

where $n$ is the number of critical or non-critical applications.
In all experiments a hardware architecture consisting of ten ECUs is used. 
Each ECU has a capacity of $\gls{symb:S_e} = 175$ slots that can be allocated and reserved by critical and non-critical tasks resulting in $\gls{symb:S_t} = 1750$ slots that are overall supplied by the architecture.
In each setup the overall amount of applications $\gls{symb:N_a}$ mapped onto the architecture was set to $40$ applications consisting of five tasks each.
In the experiments the amount of critical $\gls{symb:N_c}$ and non-critical $\gls{symb:N_{nc}}$ applications is varied to evaluate scenarios where resources are sufficiently available but also scenarios where resources are limited. 

\subsection{Graceful Degradation vs. State-of-the-Art}
\label{experiment_default}

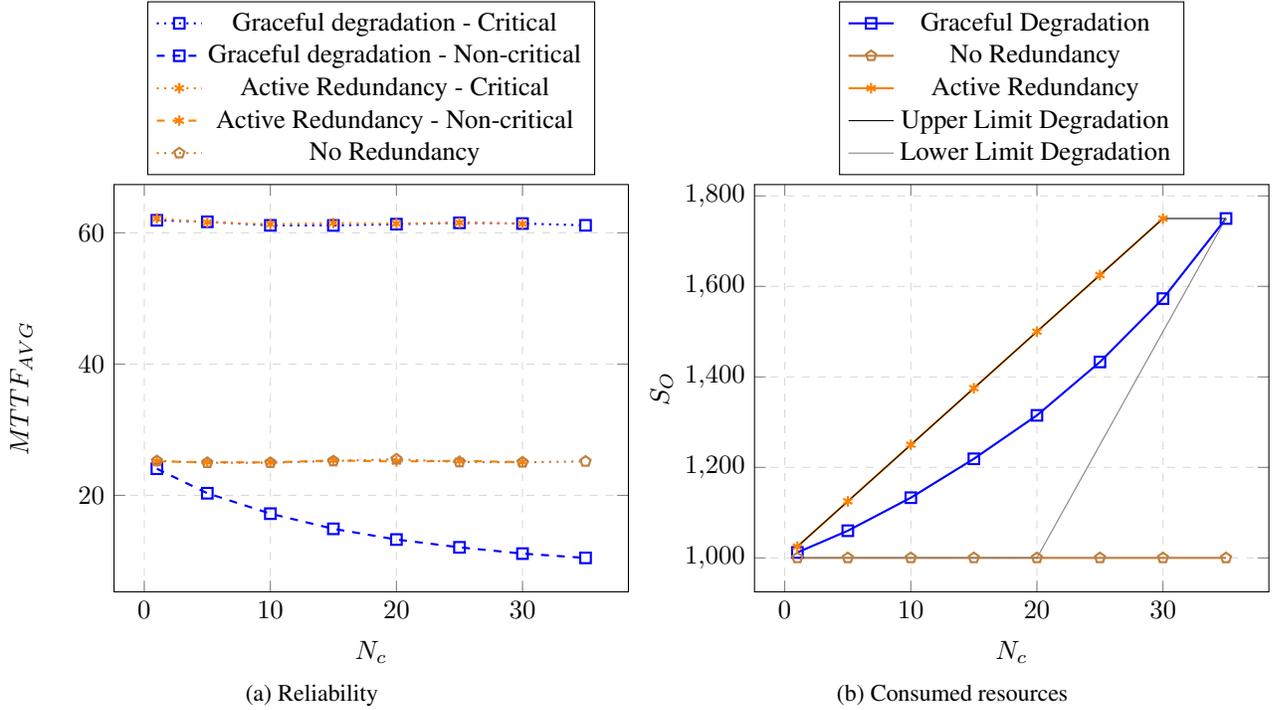
\begin{figure*}[ht]
\subfloat[Reliability \label{fig_MTTF_default}]{
	\begin{tikzpicture}
	\begin{axis}[
		width=0.51\linewidth, 
		grid=major, 
		height=7cm,
		grid style={dashed,gray!30},
		xlabel=$\gls{symb:N_c}$, 
		ylabel=$\gls{symb:MTTF_{AVG}}$,
		legend style={at={(0.5,1.45)},anchor=north},
		]
		\addplot 
		[blue, mark = square, mark options={solid}, dotted, thick] table [x = {Number}, y = {R_MTTF_C_R}, col sep = semicolon] {CSV/AliSim2.csv}; 
		\addplot 
		[blue,  mark = square, mark options={solid}, dashed, thick] table [x = {Number}, y = {R_MTTF_N_R}, col sep = semicolon] {CSV/AliSim2.csv}; 
		\addplot 
		[orange, mark = asterisk, mark options={solid}, dotted, thick] table [x = {Number}, y = {R_MTTF_C_A}, col sep = semicolon] {CSV/AliSim2.csv}; 
		\addplot 
		[orange, mark = asterisk, mark options={solid}, dashed, thick] table [x = {Number}, y = {R_MTTF_N_A}, col sep = semicolon] {CSV/AliSim2.csv}; 
		\addplot 
		[brown,  mark = pentagon, mark options={solid}, dotted, thick] table [x = {Number}, y = {NR_MTTF_C_R}, col sep = semicolon] {CSV/AliSim2.csv}; 

		\legend{Graceful degradation - Critical, 
		        Graceful degradation - Non-critical,
		        Active Redundancy - Critical,
		        Active Redundancy - Non-critical,
		        No Redundancy}
	\end{axis}
	\end{tikzpicture}
}
\hfill
\subfloat[Consumed resources \label{fig_resource_costs_default}]{
    \begin{tikzpicture}
	\begin{axis}[
	width=0.51\linewidth, 
	height=7cm,
	grid=major, 
	grid style={dashed,gray!30},
	xlabel=$\gls{symb:N_c}$, 
	ylabel= $\gls{symb:S_O}$,
	legend style={at={(0.5,1.45)},anchor=north},
	]
	\addplot 
	[blue, mark = square, thick] table [x = {Number}, y = {R_slots_R}, col sep = semicolon] {CSV/AliSim4.csv};
	\addplot 
	[brown,  mark = pentagon, thick] table [x = {Number}, y = {NR_slots_R}, col sep = semicolon] {CSV/AliSim4.csv};
	\addplot 
    [orange,  mark = asterisk, thick] table [x = {Number}, y = {Active}, col sep = semicolon] {CSV/AliSim4.csv};
	\addplot 
    [black] table [x = {Number}, y = {MAX}, col sep = semicolon] {CSV/AliSim4.csv};
    \addplot 
    [gray] table [x = {Number}, y = {Min}, col sep = semicolon] {CSV/AliSim4.csv};
	\legend{Graceful Degradation,
	        No Redundancy,
	        Active Redundancy,
	        Upper Limit Degradation,
	        Lower Limit Degradation}
	\end{axis}
	\end{tikzpicture}
}
\caption{Experimental results presenting the $\gls{symb:MTTF_{AVG}}$ and number of consumed slots $\gls{symb:S_O}$ of our graceful degradation approach (blue plot lines with square marks), an active redundancy approach (orange plot lines with asterisk marks) and no redundancy (brown plot lines with pentagram marks) scenario over an increasing number of critical application $\gls{symb:N_c}$.
Our graceful degradation approach  significantly reduces the resource consumption compared to an active redundancy approach  while guaranteeing the same reliability to critical applications. 
In this example it is also possible to fit five more critical applications  onto the same hardware platform with our graceful degradation approach than with active redundancy.
}
\end{figure*}

Subfigure \ref{fig_MTTF_default} presents the average $\gls{symb:MTTF_{AVG}}$ separated for non-critical and critical applications for a scenario with no redundancy and a scenario with our solution where redundancy and graceful degradation is applied.
In this example we used the \textit{Random} strategy for allocating and reserving slots as the default solution.
Other strategies are evaluated in the following experiments.

In the case of no redundancy (brown dotted plot line with pentagram marks), the average $\gls{symb:MTTF_{AVG}}$ is constant over all experiments for both critical and non-critical applications. 
With an active redundancy approach (orange plot lines with asterisk marks) it can be observed that the $\gls{symb:MTTF_{AVG}}$ more than doubles for critical applications compared to no redundancy while the value stays constant for non-critical applications.
However, it was not possible to evaluate a scenario with $\gls{symb:N_c} = 35$ critical applications here as the resource limit was reached with $\gls{symb:N_c} = 30$ critical applications.
Using our approach with passive redundancy and graceful degradation (blue plot lines with square marks) the same $\gls{symb:MTTF_{AVG}}$ for critical applications as with an active redundancy approach could be reached.
As the approach is more resource-saving than active redundancy it is also possible to fit $\gls{symb:N_c} = 35$ critical applications on the same hardware platform.
In the case of non-critical applications a steady decline of the $\gls{symb:MTTF_{AVG}}$ with an increasing amount of critical applications can be observed. 
With more critical applications on the same hardware architecture resources become more limited leading to an increased chance that slots that are allocated by non-critical tasks are also reserved by critical tasks.
\markasnew{This means there are more possible scenarios where in a failure scenario non-critical applications are shut down to save a critical application resulting in a reduced reliability for non-critical applications compared to the cases of no redundancy and active redundancy.}

This becomes more obvious when looking at Subfigure \ref{fig_resource_costs_default} which presents the resource consumption as the total number of occupied slots $\gls{symb:S_O}$ of all applications.
Next to the three scenarios graceful degradation, active redundancy, and no redundancy, the plot also shows the analytically derived lower and upper limit for the resource consumption of our graceful degradation solution.
The resource consumption of all applications without any redundancy stays constant at $\gls{symb:S_O} = 1000$ slots.
The resource consumption for active redundancy follows the upper limit until the maximum available resource of $\gls{symb:S_O} = 1750$ slots on the platform are hit at $\gls{symb:N_c} = 30$ critical applications.
The resource consumption of our graceful degradation solution increases steadily with an increasing amount of critical applications reaching the maximum amount of $\gls{symb:S_O} = 1750$ possible slots, which is the maximum capacity of the hardware architecture, at $\gls{symb:N_c} = 35$. 
If the number of critical applications would be increased further not all applications could fit onto the platform.
It can be observed that our solution is located roughly in the middle between the upper and the lower limit.

\markasnew{
To further compare our graceful degradation approach with active redundancy we introduce two metrics. 
We define }
\begin{equation}
    \gls{symb:MTTF_{reduction,nc}}= - \frac{\gls{symb:MTTF_{AVG,active,nc}}-\gls{symb:MTTF_{AVG,deg,nc}}}{\gls{symb:MTTF_{AVG,active,nc}}},
\end{equation}
\markasnew{
as the percental MTTF reduction of non-critical applications of our degradation approach compared to the active redundancy approach with $\gls{symb:MTTF_{AVG,active,nc}}$ representing the average MTTF of non-critical applications for the active redundancy approach and $\gls{symb:MTTF_{AVG,deg,nc}}$ representing the average MTTF of non-critical applications for our graceful degradation approach.
For the resource consumption we define $\gls{symb:S_{OH,deg}}$ as the slot overhead introduced by the degradation approach as $\gls{symb:S_{OH,deg}} = \gls{symb:S_{O,deg}} - \gls{symb:S_{O,no}}$, with $\gls{symb:S_{O,deg}}$ being the total number of consumed slots of the degradation approach and $\gls{symb:S_{O,no}}$ being the total number of consumed slots when using no redundancy. 
Furthermore, we define $\gls{symb:S_{OH,active}} = \gls{symb:S_{O,active}}- \gls{symb:S_{O,no}}$ as the slot overhead introduced by the active redundancy approach, with $\gls{symb:S_{O,active}}$ being the total number of consumed slots of the active redundancy approach.
Now we can define 
}

\begin{equation}
    \gls{symb:R_{savings}} = \frac{\gls{symb:S_{OH,active}}-\gls{symb:S_{OH,deg}}}{\gls{symb:S_{OH,active}}},
\end{equation}

\markasnew{
as the percental resource savings of our degradation approach over the active redundancy approach.
}
\markasnew{Overall the advantage of our graceful degradation approach becomes visible. 
Our approach consumes significantly less resources than an active redundancy approach while still being able to maintain the same MTTF for critical applications.
In this example, it is also possible to map five more critical applications onto the same architecture before the resource capacity is reached.
This advantage is bought by a decreased reliability of non-critical applications.
}

\begin{figure*}[ht]{
	\begin{tikzpicture}
	\begin{axis}[
		width=0.51\linewidth, 
		grid=major, 
		height=5cm,
		grid style={dashed,gray!30},
		xlabel=$\gls{symb:N_c}$, 
		ylabel= $\%$,
		legend style={at={(0.5,1.45)},anchor=north},
		xmin=-2,xmax=35,
		]
		\addplot 
		[olive,  mark = square, mark options={solid}, dashed, thick] table [x = {Number}, y = {R_savings}, col sep = semicolon] {CSV/AliSim4_savings.csv}; 
		\addplot 
		[cyan, mark = asterisk, mark options={solid}, solid, thick] table [x = {Number}, y = {MTTF_reduction}, col sep = semicolon] {CSV/AliSim2_savings.csv}; 
		\addplot 
		[black, domain=-3:40, solid, thick, samples=2] {0}; 

		\legend{$\gls{symb:R_{savings}}$, 
		        $\gls{symb:MTTF_{reduction,nc}}$
		        }
	\end{axis}
	\end{tikzpicture}
}

\caption{\markasnew{Experimental results presenting the resource savings $\gls{symb:R_{savings}}$ (dashed olive plot line with square marks) and the MTTF reduction $\gls{symb:MTTF_{reduction,nc}}$ of the non-critical applications(solid cyan plot line with asterisk marks) of our graceful degradation approach compared to an active redundancy approach.}
}
\label{fig_savings}
\end{figure*}
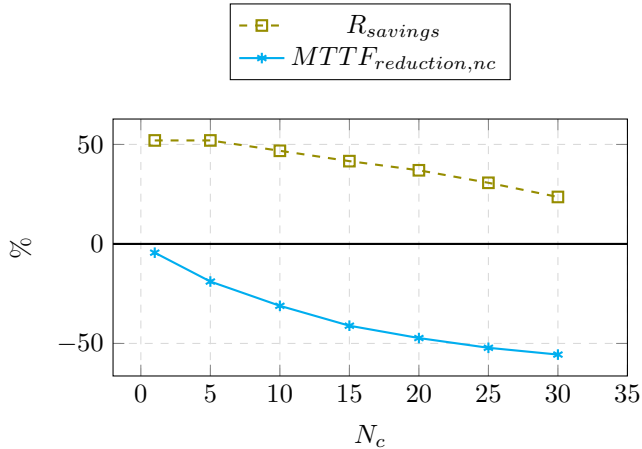

Figure \ref{fig_savings} \markasnew{presents our two metrics over the data points obtained through our experiments. 
It can be observed that the percental resource savings $\gls{symb:R_{savings}}$ of our degradation approach declines with an increasing number of critical applications. 
The potential for percental cost savings decreases as with more critical applications and a more constrained resource situation less slots for allocation from non-critical applications become available. 
This leads to a higher allocation of non-occupied slots and adds more resource overhead. 
The $\gls{symb:MTTF_{reduction,nc}}$ sinks over time until it reaches a a minimum of $-55.6\%$ at $\gls{symb:N_c}=30$. 
With an increasing number of critical applications more slots that are allocated by non-critical applications are also reserved by critical applications leading to the reduction.}

Consequently, our graceful degradation methodology can significantly reduce the resource consumption and costs if a decreased reliability of non-critical applications can be tolerated while guaranteeing the same reliability to critical applications as active redundancy approaches.

\subsection{Allocation and Reservation Strategies}
\label{experiments_strategies}

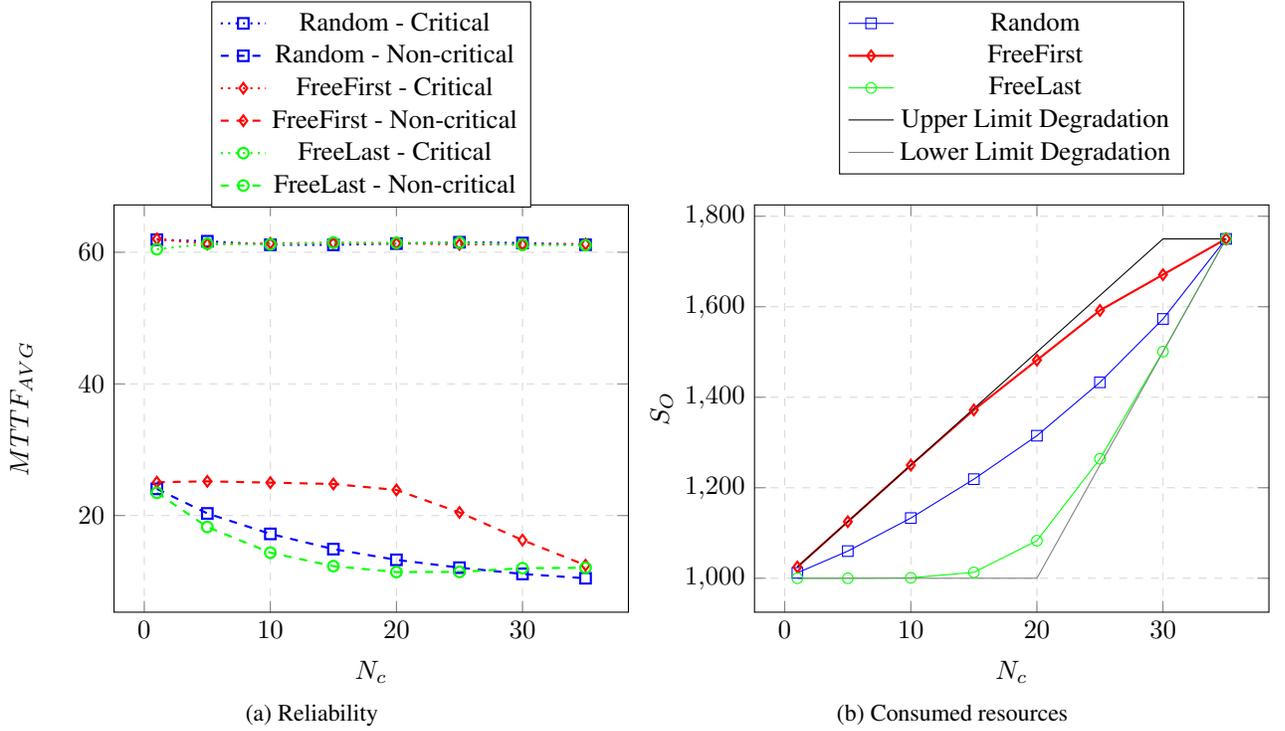
\begin{figure*}[ht]
\subfloat[Reliability \label{fig_MTTF_strategies}]{
    \begin{tikzpicture}
	\begin{axis}[
		width=0.51\linewidth, 
		grid=major, 
		height=7cm,
		grid style={dashed,gray!30},
		xlabel=$\gls{symb:N_c}$, 
		ylabel=$\gls{symb:MTTF_{AVG}}$,
		legend style={at={(0.5,1.5)},anchor=north},
		]
		\addplot 
		[blue,  mark = square, mark options={solid}, dotted, thick] table [x = {Number}, y = {R_MTTF_C_R}, col sep = semicolon] {CSV/AliSim2.csv}; 
		\addplot 
		[blue,  mark = square, mark options={solid}, dashed, thick] table [x = {Number}, y = {R_MTTF_N_R}, col sep = semicolon] {CSV/AliSim2.csv}; 
		\addplot 
		[red,  mark = diamond, mark options={solid}, dotted, thick] table [x = {Number}, y = {R_MTTF_C_FF}, col sep = semicolon] {CSV/AliSim2.csv}; 
		\addplot 
		[red,  mark = diamond, mark options={solid}, dashed, thick] table [x = {Number}, y = {R_MTTF_N_FF}, col sep = semicolon] {CSV/AliSim2.csv}; 
		\addplot 
		[green,  mark = o, mark options={solid}, dotted, thick] table [x = {Number}, y = {R_MTTF_C_FL}, col sep = semicolon] {CSV/AliSim2.csv}; 
		\addplot 
		[green,  mark = o, mark options={solid}, dashed, thick] table [x = {Number}, y = {R_MTTF_N_FL}, col sep = semicolon] {CSV/AliSim2.csv};
		\legend{Random - Critical, 
		        Random - Non-critical,
		        FreeFirst - Critical,
		        FreeFirst - Non-critical,
		        FreeLast - Critical,
		        FreeLast - Non-critical}
	\end{axis}
	\end{tikzpicture}
}
\hfill
\subfloat[Consumed resources \label{fig_resource_costs_strategies}]{
	\begin{tikzpicture}
	\begin{axis}[
	width=0.51\linewidth, 
	height=7cm,
	grid=major, 
	grid style={dashed,gray!30},
	xlabel=$\gls{symb:N_c}$, 
	ylabel= $\gls{symb:S_O}$,
	legend style={at={(0.5,1.5)},anchor=north},
	]
	\addplot 
	[blue, mark = square] table [x = {Number}, y = {R_slots_R}, col sep = semicolon] {CSV/AliSim4.csv}; 
	\addplot 
	[red,  mark = diamond, thick] table [x = {Number}, y = {R_slots_FF}, col sep = semicolon] {CSV/AliSim4.csv}; 
	\addplot 
	[green,  mark = o] table [x = {Number}, y = {R_slots_FL}, col sep = semicolon] {CSV/AliSim4.csv};
	\addplot 
    [black] table [x = {Number}, y = {MAX}, col sep = semicolon] {CSV/AliSim4.csv};
    \addplot 
    [gray] table [x = {Number}, y = {Min}, col sep = semicolon] {CSV/AliSim4.csv};
	\legend{Random,
	        FreeFirst , 
	        FreeLast, 
	        Upper Limit Degradation,
	        Lower Limit Degradation}
	\end{axis}
	\end{tikzpicture}
}
\caption{
Experimental results presenting the $\gls{symb:MTTF_{AVG}}$ and number of consumed slots $\gls{symb:S_O}$ of our three allocation and reservation strategies \textit{Random} (blue plot lines with square marks), \textit{FreeFirst} (red plot lines with diamond marks) and \textit{FreeLast} (green plot lines with circle marks) over an increasing number of critical application $\gls{symb:N_c}$.
The \textit{FreeFirst} strategy maximizes the use of available resources in order to reduce the degradation effect on non-critical applcations but also increases the resource consumption.
By contrast, the \textit{FreeLast} strategy uses the resources most efficiently reducing the consumption of hardware resources at the cost of a reduced reliability of non-critical applications.
}
\end{figure*}

In this subsection we are evaluating and comparing the MTTF and resource costs of the three allocation and reservation strategies \textit{Random}, \textit{Free-Last} and \textit{Free-First} which were introduced in Subsection \ref{section_strategies}.
Subfigure \ref{fig_MTTF_strategies} presents the MTTF for all three strategies separated into critical and non-critical applications.
The \textit{Random} strategy has already been evaluated in the previous experiment in Subsection \ref{experiment_default}.

It can be observed that the allocation and reservation strategy does not change the average MTTF of critical applications (dotted plot lines).
However, the plot lines for the MTTF of the non-critical application differ from each other.
The \textit{Free-First} strategy (red dashed plot line with diamond shapes) keeps a significantly higher MTTF with an increasing number of critical applications and starts dropping later in the plot compared to the other two strategies.
By contrast, the MTTF of \textit{Free-Last} strategy (green dashed plot line with circle shapes) drops even earlier than with the \textit{Random} strategy (blue dashed plot line with square shapes).
The \textit{Free-First} strategy allocates and reserves first any slots that have not been occupied and, therefore, reduces the number of overlapping slots and by that the degradation impact on non-critical applications. 
However, as soon as resources become more limited, finding free slots becomes more difficult resulting in a reduced MTTF for non-critical applications.
On the opposite the \textit{Free-Last} strategy tries to achieve a maximum overlap of slots by choosing occupied slots first and only allocates or reserves free slots if not otherwise possible.
This also results in a maximum possible degradation effect on non-critical applications and on average a lower MTTF.

Subfigure \ref{fig_resource_costs_strategies} presents the resource costs of the three strategies as the total number of slots occupied by all applications on the platform.
The plot line of the \textit{Free-First} strategy (red plot line with diamond shape) stays at the upper limit while the plot line of the \textit{Free-Last} strategy stays at the lower limit (green plot line with circle shape).

These results confirm that \textit{Free-First} strategy maximizes the use of the available resources in order to reduce the degradation effect, but might increase the resources costs.
Therefore, for a given amount of resources the strategy avoids unnecessary degradation by using up the resources as far as possible.
By contrast, the \textit{Free-Last} strategy uses the resources most efficiently reducing the consumption of hardware resources at the cost of a reduced reliability of non-critical applications.
Overall these experiments show that \textit{Free-First} strategy minimizes the degradation effect while \textit{Free-Last} strategy maximizes it.

\subsection{Exposure Reduction}
\label{experiments_exposure}

\begin{figure}[ht]
	\begin{center}
		\begin{tikzpicture}
		\begin{axis}[
			width=\linewidth, 
			height=8cm,
			grid=major, 
			grid style={dashed,gray!30},
			xlabel=$N_{c}$, 
			ylabel=$\gls{symb:MTTF_{AVG}}$,
			legend style={at={(0.5,1.5)},anchor=north},
			]
			\addplot 
			[blue,  mark = square, mark options={solid}, dotted, thick] table [x = {Number}, y = {R_MTTF_C_R}, col sep = semicolon] {CSV/AliSim5.csv}; 
			\addplot 
			[blue,  mark = square, mark options={solid}, dashed, thick] table [x = {Number}, y = {R_MTTF_N_R}, col sep = semicolon] {CSV/AliSim5.csv}; 
			\addplot 
			[red,  mark = diamond, mark options={solid}, dotted, thick] table [x = {Number}, y = {R_MTTF_C_FF}, col sep = semicolon] {CSV/AliSim5.csv}; 
			\addplot 
			[red,  mark = diamond, mark options={solid}, dashed, thick] table [x = {Number}, y = {R_MTTF_N_FF}, col sep = semicolon] {CSV/AliSim5.csv}; 
			\addplot 
			[green,  mark = o, mark options={solid}, dotted, thick] table [x = {Number}, y = {R_MTTF_C_FL}, col sep = semicolon] {CSV/AliSim5.csv}; 
			\addplot 
			[green,  mark = o, mark options={solid}, dashed, thick] table [x = {Number}, y = {R_MTTF_N_FL}, col sep = semicolon] {CSV/AliSim5.csv};
			\addplot 
			[brown,  mark = pentagon, thick] table [x = {Number}, y = {NR_MTTF_C_R}, col sep = semicolon] {CSV/AliSim5.csv}; 
			\legend{Random - Critical, 
    		        Random - Non-critical,
    		        FreeFirst - Critical,
    		        FreeFirst - Non-critical,
    		        FreeLast - Critical,
    		        FreeLast - Non-critical,
			        No Redundancy}
		\end{axis}
		\end{tikzpicture}
		\caption{
		Experimental results presenting the $\gls{symb:MTTF_{AVG}}$ of our three allocation and reservation strategies \textit{Random} (blue plot lines with square marks), \textit{FreeFirst} (red plot lines with diamond marks) and \textit{FreeLast} (green plot lines with circle marks) using our \textit{Predeccesor Heuristic} over an increasing number of critical application $\gls{symb:N_c}$.
		Compared to previous experiments, the $\gls{symb:MTTF_{AVG}}$ of both critical and non-critical applications increased significantly as the exposure of applications to different failure sources is reduced.
		The $\gls{symb:MTTF_{AVG}}$ decreases with an increasing number of critical applications $\gls{symb:N_c}$ as resources become more limited and the chance to map all task instances on the same two ECUs decreases, distributing the critical application more on the hardware platform.
		}
	\label{fig_ecu_heuristic}
	\end{center}
\end{figure}
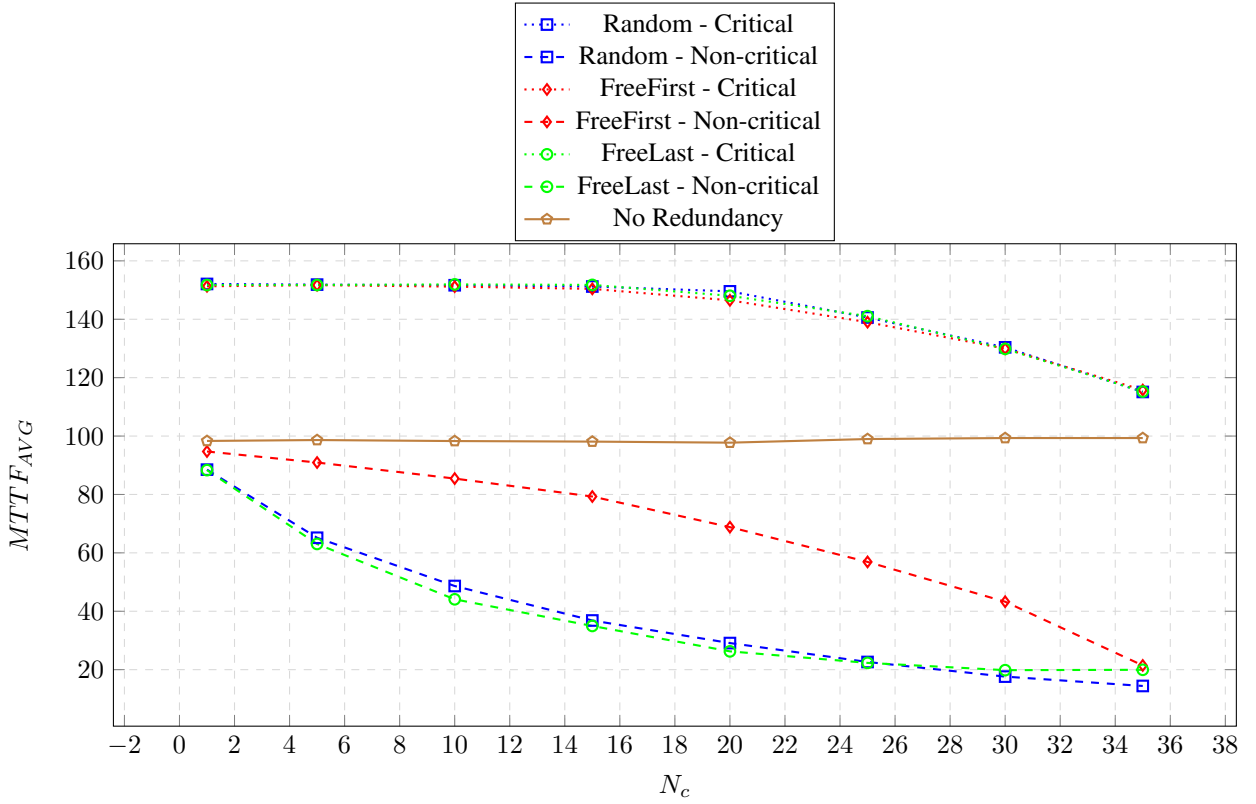

In previous experiments the ECU of a task instance was being chosen randomly by the agent and the task was mapped to the next ECU with available resources. 
This led to scenarios were both critical and non-critical applications were highly distributed over the whole hardware platform. 
In terms of reliability this would mean that more ECU failures could lead to the failure of an application. 
However, it would be preferable in terms of reliability to reduce the exposure to different failure sources.
Therefore, we introduce the \textit{Predecessor Heuristic} where agents do not choose ECUs randomly but prefer ECUs to which preceding task instances are already mapped to concentrating the task instances on a smaller number of ECUs.

Figure \ref{fig_ecu_heuristic} presents the results for the three allocation and reservation strategies \textit{Random}, \textit{FreeFirst} and \textit{FreeLast} as well the resulst for the case with no redundancy.
Comparing the scenario without redundancy (brown plot line with pentagon marks) to Subfigure \ref{fig_MTTF_default}, the MTTF increased from around $25$ to almost $100$.
Here, the tasks of one application are on average almost always mapped to only on ECU, while previously applications were scattered around $4.1$ ECUs on average.
The MTTF is even exceeding the MTTF of critical applications from Subfigure \ref{fig_MTTF_default}.
Even though the critical applications had a double redundancy, they were distributed around $6.5$ ECUs on average which increased the exposure to failure significantly.

In Figure \ref{fig_ecu_heuristic} the plot lines for the critical applications of all three strategies (dotted plot lines) are similar.
The MTFF starts around $152$ until it hits a low around an MTTF of $115$ at $\gls{symb:N_c} = 35$.
Initially, agents of critical applications manage to distribute the task instances mostly on two ECUs (at least two ECUs have to be involved to ensure double redundancy).
However, as resources become more limited, the chance to map all task instances on the same two ECUs decreases, distributing the critical application more on the hardware platform.
This effect is not visible in Subfigure \ref{fig_MTTF_default} as the ECUs were already chosen randomly.

The general course of the plot lines of non-critical applications of all three strategies (dashed plot lines) is similar as in Subfigure \ref{fig_MTTF_strategies} but there are a few differences.
Overall the MTTF starts at slighlty below 100 until it hits a low around an MTTF of 20 at $\gls{symb:N_c} = 35$. 
Here, the degradation has much larger impact on the MTTF than in the previous example.
The reason is that the MTTF is starting with a high value were applications are mostly mapped to only one ECU and, thus, only the failure of one ECU can cause an application failure.
When resource become more limited, more slots of non-critical task are getting reserved from many different critical applications, which themselves are also getting more distributed on the platform.
Overall this increases the exposure of non-critical applications to about $5$ ECUs on average.
This experiment has shown that while redundancy can significantly increase reliability, reducing the exposure of applications to different failure sources is a leverage that should not be underestimated.

\subsection{Summary}
\label{summary_experiments}

Overall our experiments have shown that graceful degradation significantly reduces resource consumption while still maintaining the same reliability as an active redundancy approach for critical applications.
Furthermore, it is possible to fit more applications onto the same hardware platform as resources are being used more efficiently.
This advantage is bought with a decreased reliability of non-critical applications.
Additionally, we proposed and evaluated the three allocation and reservation strategies \textit{Random}, \textit{FreeFirst} and \textit{FreeLast}. 
The \textit{FreeFirst} strategy maximizes the use of available resources and reduces the degradation effect on non-critical applications but also increases the resource consumption.
By contrast, the \textit{FreeLast} strategy minimizes the resource consumption at the cost of a reduced reliability of non-critical applications.
Last we evaluated our \textit{Predecessor Heuristic} where tasks preferred ECUs to which already other task of the same application were mapped to. 
Here, the reliability of both critical and non-critical applications increased significantly as the exposure of applications to different failure sources was reduced.
Summarized, graceful degradation can be a powerful methodology
which uses resources more efficiently than common redundancy approaches and which can strongly increase the number of applications that can be mapped onto the same system architecture while providing the same fail-operational
capabilities.

\section{Conclusion}
In this paper we presented a reliability analysis of gracefully degrading automotive systems. 
We introduced our graceful degradation approach which is based on composable scheduling such that tasks can independently allocate and reserve resources.
Furthermore, we described our agent-based mapping approach which finds application mapping at run-time. 
In a failure scenario, once a critical passive task has to be started, it can take over the resources such that the system is being degraded in an intended way.
Here, we proposed the three allocation and reservation strategies \textit{Random}, \textit{FreeFirst}, and \textit{FreeLast}, which heavily influence how the system will be degraded.
With a graceful degradation approach, the reliability of critical applications is increased at the cost of a reliability decrease of non-critical applications.
To quantify and understand the effect that a graceful degradation approach has on both critical and non-critical applications we gave a short overview over state-of-the-art reliability analysis and presented our approach to formally analyze the impact of graceful degradation on the reliability of critical and non-critical applications.
We performed multiple experiments on our in-house developed simulation platform to evaluate our graceful degradation approach.
Compared to active redundancy, graceful degradation can significantly reduce resource consumption while still maintaining the same reliability to critical applications.
However, this advantage is bought with a reduced reliability of non-critical applications.
In resource-constrained scenarios, the graceful degradation approach is able to map more applications on the same hardware platform than with an active redundancy approach.
Our experiments also showed that the \textit{FreeFirst} strategy maximizes the use of available resources and reduces the degradation effect on non-critical applications but also increases the resource consumption.
By contrast, the \textit{FreeLast} strategy minimizes the resource consumption at the cost of a reduced reliability of non-critical applications.
The experimental results with the \textit{Predecessor Heuristic} showed that the reliability of critical and non-critical applications could be increased significantly as the exposure of applications to different failure sources was reduced.
Summarized, our results confirm that graceful degradation can be a powerful methodology which significantly reduces resource consumption compared to active redundancy while providing the same fail-operational capabilities to critical applications.

\bibliographystyle{unsrtnat}


\printglossary[type=\acronymtype, nonumberlist]
\printglossary[type=reliability, nonumberlist]

\end{document}